\documentclass[acmsmall, screen]{acmart}
\acmJournal{TOSEM}

\usepackage{subcaption}
\usepackage{fancyvrb}
\usepackage{xurl}
\usepackage{booktabs}
\usepackage{url}
\usepackage{amsmath,amsfonts}
\usepackage{algorithmic}
\usepackage{xspace}
\usepackage{tabularx}
\usepackage{amsmath}
\usepackage{balance}

\usepackage{bm}

\usepackage[frozencache,cachedir=.]{minted}

\definecolor{myhighlight}{rgb}{0.9, 0.9, 0.5}
\setminted[java]{frame=lines,linenos,breaklines,xleftmargin=40pt,xrightmargin=30pt,fontsize=\footnotesize,highlightcolor=myhighlight}
\usepackage[many]{tcolorbox}    	
\definecolor{darkgreen}{rgb}{0, 0.5, 0} 
\definecolor{whitesmoke}{rgb}{0.99, 0.99, 0.99} 
\definecolor{main}{HTML}{D0D3D4}    
\definecolor{sub}{HTML}{D0D3D4}     
\tcbset{
    sharp corners,
    colback = whitesmoke,
    before skip = 0.2cm,  
    after skip = 0.1cm,      
    boxsep=2pt,
    left=5pt,
    right=5pt,
    top=5pt,
    bottom=5pt
}

\newcommand{\rqa}{$RQ_1$}
\newcommand{\rqb}{$RQ_2$}
\newcommand{\rqc}{$RQ_3$}
\newcommand{\rqd}{$RQ_4$}
\newcommand{\rqaa}{How prevalent is SATD in test code?}
\newcommand{\rqbb}{Is SATD in test code correlated with test quality?}
\newcommand{\rqcc}{What are the purposes of SATD in test code?}

\newcommand{\rqdd}{To what extent can SATD in Test Code be classified automatically?}

\newcommand{\rqA}{\rqa: \rqaa}
\newcommand{\rqB}{\rqb: \rqbb}
\newcommand{\rqC}{\rqc: \rqcc}
\newcommand{\rqD}{\rqd: \rqdd}

\newcommand{\ts}{Test SATD\xspace}
\newcommand{\satd}{SATD\xspace}

\newcommand{\testsmells}{test smells\xspace}
\newcommand{\codesmells}{code smells\xspace}

\newcommand{\ProdOneQuartile}{7.3\xspace}
\newcommand{\ProdThreeQuartile}{18.9\xspace}
\newcommand{\ProdAve}{15.6\xspace}
\newcommand{\ProdMedian}{10.9\xspace}
\newcommand{\TestOneQuartile}{1.6\xspace}
\newcommand{\TestThreeQuartile}{14.1\xspace}
\newcommand{\TestAve}{9.5\xspace}
\newcommand{\TestMedian}{4.1\xspace}
\newcommand{\PerDiff}{62.4\%\xspace}
\newcommand{\EffectSize}{0.365\xspace}

\newcommand{\SampleSize}{506\xspace}
\newcommand{\NumberOfDetails}{20\xspace}
\newcommand{\Matches}{407\xspace}
\newcommand{\MatchRate}{80.4\%\xspace}
\newcommand{\Kappa}{0.78\xspace}

\newcommand{\FP}{36\xspace}
\newcommand{\UnCat}{40\xspace}
\newcommand{\ManualSampleSize}{430\xspace}

\newcommand{\FailuresEnvironment}{11\xspace}
\newcommand{\FailuresInput}{4\xspace}
\newcommand{\OnholdDefect}{42\xspace}
\newcommand{\OnholdUnimplement}{18\xspace}
\newcommand{\OnholdDebug}{7\xspace}
\newcommand{\IncomleteTest}{120\xspace}
\newcommand{\ExtraTest}{17\xspace}
\newcommand{\Workaround}{53\xspace}
\newcommand{\NonOptimal}{44\xspace}
\newcommand{\BetterWay}{37\xspace}
\newcommand{\DoubtsTest}{14\xspace}
\newcommand{\DoubtsTestImp}{9\xspace}
\newcommand{\NonFunctional}{6\xspace}
\newcommand{\AskTestUpdate}{17\xspace}
\newcommand{\AskRefactoring}{9\xspace}
\newcommand{\MoreDocument}{7\xspace}
\newcommand{\FlakyTest}{5\xspace}
\newcommand{\AskProdUpdateInTest}{4\xspace}
\newcommand{\TestDeletion}{4\xspace}
\newcommand{\UnusedTest}{2\xspace}

\newcommand{\ProdIssues}{82\xspace}
\newcommand{\TestIssues}{348\xspace}

\newcommand{\Failures}{15\xspace}
\newcommand{\OnHoldTask}{67\xspace}
\newcommand{\TestCompleteness}{137\xspace}
\newcommand{\TestImplementation}{163\xspace}
\newcommand{\TestMaintenance}{48\xspace}

\usepackage{xcolor}
\definecolor{darkgreen}{rgb}{0, 0.5, 0} 
\definecolor{whitesmoke}{rgb}{0.99, 0.99, 0.99} 

\def\Underline{\setbox0\hbox\bgroup\let\\\endUnderline}
\def\endUnderline{\vphantom{y}\egroup\smash{\underline{\box0}}\\}
\def\|{\verb|}

\newcommand{\ie}{\textit{i.e.,}\xspace}
\newcommand{\eg}{\textit{e.g.,}\xspace}

\newcommand{\etal}{\xspace\textit{et al.}\xspace}

\newcounter{findings_no}

\usepackage{listings}
\definecolor{backcolour}{rgb}{0.95,0.95,0.92}
\lstdefinelanguage{diff}{
  morecomment=**[f][\color{red}]{-},         
  morecomment=**[f][\color{darkgreen}]{+},       
  moredelim=**[is][\bfseries]{@@}{@@},
}
\definecolor{backcolour}{rgb}{0.95,0.95,0.92}
\lstdefinelanguage{commit}{ 
  breakindent = 0pt,
  numbers=none,
  backgroundcolor=\color{white},
  frame=single,
  xleftmargin=3.5em,
  numbersep=0em,
  xrightmargin=1.5em,
}

{\list{}{%
\rightmargin=7pt%
\leftmargin=7pt%
}\item\relax}{\endlist}

\usepackage{framed}
\usepackage{chngpage}
\definecolor{formalshade}{rgb}{1,1,1}

\newenvironment{formal}{%
\vspace{5pt}
  \MakeFramed{\advance\hsize-\width\FrameRestore}%
  \noindent\hspace{-4.55pt}
  \begin{adjustwidth}{}{}%
  \vspace{0pt}\vspace{0pt}%
}
{%
  \vspace{2pt}\end{adjustwidth}\endMakeFramed%
\vspace{-6pt}
}

\usepackage{multirow} 
\usepackage{listings}

\AtBeginDocument{%
  }

\setcopyright{acmlicensed}
\copyrightyear{2025}
\acmYear{2025}
\acmDOI{XXXXXXX.XXXXXXX}

\acmISBN{978-1-4503-XXXX-X/18/06}

\begin{document}

\title{Understanding Self-Admitted Technical Debt in Test Code: An Empirical Study}

\author{Ibuki Nakamura}
\email{nakamura.ibuki.nh4@naist.ac.jp}
\orcid{0009-0001-0671-7454}
\affiliation{%
  \institution{Nara Institute of Science and Technology}
  \city{Ikoma}
  \country{Japan}
}

\author{Yutaro Kashiwa}
\affiliation{%
  \institution{Nara Institute of Science and Technology}
  \city{Ikoma}
  \country{Japan}
}
\email{yutaro.kashiwa@is.naist.jp}
\orcid{0000-0002-9633-7577}

\author{Bin Lin}
\affiliation{%
  \institution{Hangzhou Dianzi University}
  \city{Hangzhou}
  \country{China}
}
\email{b.lin@hdu.edu.cn}
\orcid{0000-0001-6307-8460}

\author{Hajimu Iida}
\affiliation{%
  \institution{Nara Institute of Science and Technology}
  \city{Ikoma}
  \country{Japan}
}
\email{iida@itc.naist.jp}
\orcid{0000-0002-2919-6620}

\renewcommand{\shortauthors}{Nakamura\etal}

\begin{abstract}

Developers often opt for easier but non-optimal implementation to meet deadlines or create rapid prototypes, leading to additional effort known as technical debt to improve the code later. Oftentimes, developers explicitly document the technical debt in code comments, referred to as Self-Admitted Technical Debt (SATD). Numerous researchers have investigated the impact of SATD on different aspects of software quality and development processes. However, most of these studies focus on SATD in production code, often overlooking SATD in the test code or assuming that it shares similar characteristics with SATD in production code.
In fact, a significant amount of SATD is also present in the test code, with many instances not fitting into existing categories for the production code. This study aims to fill this gap and disclose the nature of SATD in the test code by examining its distribution and types. Moreover, the relation between its presence and test quality is also analyzed.

Our empirical study, involving 17,766 SATD comments (14,987 from production code, 2,779 from test code) collected from 50 repositories, demonstrates that while SATD widely exists in test code, it is not directly associated with test smells. Our study also presents comprehensive categories of SATD types in the test code, and machine learning models are developed to automatically classify SATD comments based on their types for easier management. 
Our results show that the CodeBERT-based model outperforms other machine learning models in terms of recall and F1-score. However, the performance varies on different types of SATD.

\end{abstract}



\received{20 February 2007}
\received[revised]{12 March 2009}
\received[accepted]{5 June 2009}


\begin{center}
\vspace*{-2cm}
\footnotesize
\textit{This manuscript is accepted at ACM Transactions on Software Engineering and Methodology (TOSEM).}
\vspace{1cm}
\end{center}

\maketitle

\section{Introduction}
\label{sec:Introduction}
Developers often opt for a simpler, albeit non-optimal, implementation over an ideal but time-consuming one due to various reasons such as meeting deadlines or creating rapid prototypes. These implementation choices may introduce additional efforts for future improvement, known as technical debt~\cite{DBLP:journals/oopsm/Cunningham93}.
While technical debt can accelerate development in the short term, it becomes a long-term obstacle, degrading software quality and impeding project progress~\cite{DBLP:journals/software/KruchtenNO12, DBLP:conf/icse/ZazworkaSSS11}. As technical debt accumulates, maintenance costs also increase, making it challenging to repay the debt~\cite{DBLP:conf/icse/NugrohoVK11, McConnell2013}.

In practice, developers frequently use keywords like \verb|TODO:| or \verb|FIXME:| in code comments to highlight issues or tasks that need to be addressed. This specific type of technical debt, intentionally introduced by developers, is known as Self-Admitted Technical Debt (SATD)~\cite{DBLP:conf/icsm/PotdarS14}. 
SATD provides opportunities for researchers and developers to identify technical debt and understand how it is handled in software projects. 
Given the impact of technical debt on the development process and product quality, researchers have done extensive work to analyze their usage in source code. 
Vassallo\etal\cite{DBLP:conf/icsm/VassalloZRBPPZ16} found that 88\% of developers working on financial systems have introduced SATD. 
In recent years, it has been reported that SATD impacts software quality~\cite{DBLP:journals/jss/SierraSK19}. 
Wehaibi\etal\cite{DBLP:conf/wcre/WehaibiSG16} compared files with and without SATD, showing that files with SATD tend to have more defects. Both Wehaibi\etal\cite{DBLP:conf/wcre/WehaibiSG16} and Kamei\etal\cite{kamei2016} observed that changes related to SATD are more complex than non-SATD changes.
Additionally, some studies have manually inspected and classified SATD to demystify it~\cite{DBLP:journals/infsof/FariasNKS20,DBLP:conf/msr/BavotaR16}. 
Farias\etal\cite{DBLP:journals/infsof/FariasNKS20} created nine categories of SATD based on technical debt definitions established by the previous studies~\cite{DBLP:journals/infsof/AlvesMNSSS16, DBLP:journals/infsof/AlvesNS18} and detected SATD using a pattern-based approach.
Furthermore, Bavota\etal\cite{DBLP:conf/msr/BavotaR16}  conducted a large empirical study and classified SATD into six categories. Based on these categories, Sala\etal\cite{DBLP:conf/ease/SalaTF21} developed DebtHunter, an ML-based tool to categorize SATD types, further facilitating SATD studies.\footnote{\url{https://github.com/PandaMinore/DebtHunter-Tool}}

However, these studies often overlook or are less eager to investigate SATD in test code. This oversight hinders test-specific issues highlighted by SATDs. For example, a comment such as, ``TODO: This test only covers the happy path - need to add edge cases for null inputs,'' reveals that while the test may pass, it provides incomplete coverage and a false sense of security, leaving the system vulnerable. Furthermore, this gap is evident in the literature; for instance, Bavota\etal's categories~\cite{DBLP:conf/msr/BavotaR16} primarily use SATD from production code, resulting in no subcategories in their ``Test'' category, while other categories have many subcategories. Sala\etal's tool~\cite{DBLP:conf/ease/SalaTF21} also automatically excludes test code when detecting SATD.

Despite this oversight in SATD research, issues in test code have garnered attention from many researchers. For example, van Deursen\etal\cite{DBLP:article/deursen/2001} introduced the concept of test smells, which represent issues in test code. These test smells indicate inadequate design or implementation, leading to maintenance difficulties and an increased defect injection rate. Additionally, Li\etal\cite{DBLP:journals/jss/LiAL15} conducted a literature review and analyzed previous studies on technical debt (non-self-admitted) and its management. Their results present several subcategories of test-related technical debt. However, the types of listed test-related technical debt are rather limited, and it remains unclear whether the types of SATD would be different compared to general technical debt in the test code.


Studying SATD in test code can help us gain comprehensive insight on how developers document test code issues. Identifying the types of SATDs in test code can help developers better manage the SATD. 
In this study, we aim to analyze the prevalence and types of SATD in test code and classify their types. The main contributions of our study are as follows. 

\begin{itemize}
    \item  We demonstrated the prevalence of SATD in test code and compared it to that in production code. Our result indicates that much fewer instances of SATD reside in the test code, but they are not negligible and still play an important role. 

    \item We examined the relationship between Test SATDs and software quality at the method level. The results reveal that test methods with SATD tend to be more lengthy and prone to code smells. However, no relationship can be found between \ts and \testsmells, implying that these issues occur independently.

    \item We conducted a manual classification of \SampleSize randomly selected Test SATD instances and identified \NumberOfDetails types of issues, which could be further categorized into 5 main groups. The most common category was ``Indicates incomplete or unimplemented tests'', highlighting the lack of proper tests in general.
    
    \item We built a CodeBERT-based model to automatically classify the types of Test SATD, which outperforms other machine learning models in terms of recall and F1-score, reaching 0.69 and 0.70, respectively. However, the performance varies on different types of SATD. 

\end{itemize}

\smallskip
\textbf{Replication Package:} 
To facilitate replication and further
studies, we provide the data and scripts used in our replication package on GitHub.\footnote{\url{https://github.com/ibu00024/UnderstandingSelfAdmittedTechnicalDebtinTestCode}}

\smallskip
{\bf Structure of this paper:}
Section \ref{sec:related} introduces related work.
Section \ref{sec:rq} presents motivating examples and the research questions we aim to address.
Section \ref{sec:Method} describes the methodology adopted to answer the research questions.
Section \ref{sec:Result} presents the empirical results for each research question.
Section \ref{sec:discussions} summarizes our findings, lessons learned, and describes the implications. 
Section \ref{sec:Threat} discusses the threats to validity and Section \ref{sec:Conclusion} concludes this paper.

\section{Related work}
\label{sec:related}
In this section, we present related work on self-admitted technical debt, test code issues, and their intersections. 

\subsection{Self-admitted Technical Debt}
Potdar and Shihab~\cite{DBLP:conf/icsm/PotdarS14} investigated technical debt explicitly documented as comments in the source code by developers, known as Self-Admitted Technical Debt (SATD). SATD comments often describe issues or incomplete tasks~\cite{DBLP:journals/tse/MaldonadoST17}, such as \textit{``TODO: - This method is too complex, lets break it up''}\footnote{\url{https://github.com/argouml-tigris-org/argouml/blob/VERSION_0_34/src/argouml-app/src/org/argouml/notation/providers/uml/AbstractMessageNotationUml.java\#L448}} in the ``ArgoUml'' project and \textit{``TODO no methods yet for getClassname''}\footnote{\url{https://github.com/apache/ant/blob/rel/1.7.0/src/main/org/apache/tools/ant/util/ClasspathUtils.java\#L496}} in the project ``Apache Ant''.
To understand the role SATD plays in software projects, numerous studies have proposed approaches to identify SATD in software and analyzed its relations with software quality. 

\subsubsection{SATD Identification} 

Various approaches have been proposed to detect SATD in code comments, which are based on either pattern-matching or machine learning. 

As one of the earliest approaches, 
Potdar and Shihab~\cite{DBLP:conf/icsm/PotdarS14} extracted comments from source code using srcML~\cite{DBLP:conf/scam/CollardDM11} and identified 62 recurring patterns representing SATD. These patterns are further used to match code comments to detect SATD. Maldonado\etal\cite{DBLP:conf/icsm/MaldonadoS15} extended this work and developed filtering heuristics to remove the comments which are less likely to contain SATD, such as license comments and commented source code. They also manually classified SATDs into five types. Farias\etal\cite{DBLP:conf/icsm/FariasNSS15} proposed a Contextualized Vocabulary
Model to detect SATD, which systematically takes into account how terms may
be combined to identify different types of debt.
Guo\etal\cite{DBLP:journals/corr/abs-1910-13238} proposed MAT (Matches Task Annotation Tags), an approach for detecting SATD based on tags such as \texttt{TODO} and \texttt{FIXME}.

Machine learning has also been adopted to detect SATD in recent years. 
For example, Maldonado\etal\cite{DBLP:journals/tse/MaldonadoST17} proposed an NLP-based method, which trains a maximum entropy classifier with a classified dataset to automatically detect SATDs related to design and requirements. Liu\etal\cite{DBLP:conf/icse/LiuHXSLL18} developed SATD Detector, which is based on a composite classifier of several sub-classifiers using Naive Bayes Multinomial (NBM). Their tool provides an environment for developers to manage SATD comments through an Eclipse plug-in. Ren\etal\cite{DBLP:journals/tosem/RenXXLWG19} proposed a method to detect SATD using convolutional neural networks (CNN). Sala\etal\cite{DBLP:conf/ease/SalaTF21} developed ``DebtHunter,'' a tool that first identifies SATD and then classifies SATD into specific debt types. Both classifiers used in these two steps are built on Sequential Minimal Optimization (SMO).

\subsubsection{Prevalence of technical debt} 
Potdar and Shihab~\cite{DBLP:conf/icsm/PotdarS14} analyzed the quantity and reasons for introducing SATD in four large-scale open-source software repositories. They found that SATD was present in 2.4\% to 31\% of the files in the projects, with only 26.3\% to 63.5\% of the issues being resolved after introduction. They also discovered that SATD is often introduced by experienced developers and tends to be added regardless of the release timing. 
Maldonado and Shihab~\cite{DBLP:conf/icsm/MaldonadoS15} investigated SATD in five open-source software repositories and classified SATD into five categories as indicated by Alves\etal\cite{DBLP:conf/icsm/AlvesRCMS14}. Among these types, Design Debt, which indicates problems that violate design principles, accounts for the largest proportion of cases (ranging from 42\% to 84\%), followed by Requirement Debt, which indicates problems related to requirements (ranging from 5\% to 45\%). Also, Vassallo\etal\cite{DBLP:conf/icsm/VassalloZRBPPZ16} conducted a survey on testing activities and the management of technical debt in financial systems. They found that many developers documented technical debt in a self-reported manner, with 88\% of developers implementing SATD. 

\subsubsection{Relation between SATD and software quality} 

Wehaibi\etal\cite{DBLP:conf/wcre/WehaibiSG16} compared the number of defects in files with and without SATDs in five OSS projects and found no clear difference. However, they observed that the number of defects tended to increase after SATDs were introduced.
Additionally, while future defects were less likely to occur after changes involving SATD, these changes were usually more complex than changes to files without SATD. Based on this finding, they concluded that the existence of SATD makes system changes more difficult. 
Kamei\etal\cite{kamei2016} investigated the complexity and dependency of code changed during the introduction and removal of SATD to measure the amount of ``interest'' it accumulated. Based on the Apache JMeter project, they found that 42\% to 44\% of the code incurred positive interest, meaning it would cost more to remove in the future. Conversely, 8\% to 13\% incurred negative interest, and 42\% to 49\% did not incur interest.


Palomba\etal\cite{DBLP:conf/iwpc/PalombaZOL17} studied the relationship between SATD and refactoring in three OSS projects. They found that approximately 46\% of refactored classes had SATD in previous versions. Furthermore, in 67\% of cases, comments and descriptions related to SATD were deleted by refactoring, resolving the technical debt. This study shows that SATD might play an important role as a motivation for refactoring. 
Maldonado\etal\cite{DBLP:journals/tse/MaldonadoST17} analyzed the relationship between SATD and code smells. Code smells indicate potential flaws in the design or implementation of production code~\cite{DBLP:books/daglib/0019908}, which are likely to cause problems in the future. 
The authors analyzed the overlap between SATD and three representative types of code smells: Long Method, God Class, and Feature Envy at the file level. Their investigation, which targeted 10 OSS repositories, revealed that 65\% of files containing SATD also had Long Methods, 44.2\% contained God Classes, and 20.7\% contained Feature Envy classes.
Moreover, when considering all code smells, they found that an average of 69.7\% of files containing SATD had at least one code smell. From these findings, they concluded that while there are certain overlaps between SATD and code smells, these two indicators can serve as complementary approaches for detecting technical debt.   


Rantala\etal\cite{DBLP:journals/sqj/RantalaML24} investigated the relationship between keyword-labeled self-admitted technical debt (KL-SATD) \cite{DBLP:conf/euromicro/RantalaM020}, such as ``TODO'' or ``FIXME'', and issues identified by static code analysis using SonarQube. Their results indicate that KL-SATD is associated with reduced code maintainability as measured by SonarQube. The introduction and removal of KL-SATD are mainly related to code smells rather than vulnerabilities or bugs. However, there is a limited overlap between KL-SATD and SonarQube issues, with only 36\% of KL-SATD comments being in the context of a SonarQube issue, and only 15\% directly addressing an issue.

These studies do not distinguish whether the SATD is present in the production code or the test code. 

\subsection{Issues in Test code}
Van Deursen\etal \cite{DBLP:article/deursen/2001} coined the concept of test smells, representing potential problems in test code, which are often used to investigate their impact on software quality and maintainability \cite{DBLP:journals/jss/GarousiK18}. Bavota\etal \cite{DBLP:journals/ese/BavotaQOLB15} investigated the prevalence of test smells in test code and their effect on software understanding and maintenance. They found that 86\% of the test cases for 27 software systems contained at least one test smell, and the presence of test smells reduced the efficiency of program comprehension. Tufano\etal \cite{DBLP:conf/kbse/TufanoPBPOLP16} conducted an empirical study involving 152 projects to investigate the nature of test smells. Their results revealed that test smells tend to persist in the system for a long time, with 80\% not being fixed even after 1,000 days. Spadini\etal \cite{DBLP:conf/icsm/SpadiniPZBB18} studied the impact of test smells on the quality of test code. Their analysis of 10 OSS projects revealed that tests with test smells are 47\% more likely to be modified and 81\% more likely to have defects than code without test smells.

These aforementioned studies have been conducted to clarify test-code-related issues that are not admitted by developers. These studies explore different types of test issues. Studying SATD in test code could provide new insights into what issues developers face in test code and help to identify new types of test smells and (non-self-admitted) technical debt. 

\subsection{Testing-Related SATD}

Gat and Heintz \cite{DBLP:conf/icse/GatH11} investigated the impact of technical debt on software development processes and proposed methods for its reduction. They found that the lack of automated testing is a major factor causing delays in development speed and recommended enhancing unit testing and utilizing ``Test as a Service.'' Codabux and Williams \cite{DBLP:conf/icse/CodabuxW12} also examined the challenges and best practices in managing technical debt. They identified ``Automation debt,'' caused by a lack of test automation, and ``Test debt,'' arising from unexecuted tests.

Li\etal\cite{DBLP:journals/jss/LiAL15} conducted a literature survey on technical debt published between 1992 and 2013. They categorized technical debt into ten different categories, identifying one relevant to testing, known as Test Technical Debt, which is the third most well-studied category. The test category includes seven subcategories: ``Low code coverage,'' ``Deferring testing,'' ``Lack of tests,'' ``Lack of test automation,'' ``Residual defects not found in tests,'' ``Expensive tests,'' and ``Estimation errors in test effort planning.''

These identified test-related SATD types are either over coarse-grained or not extracted in a systematic manner (\ie by aggregating multiple independent studies).  


\section{Research Questions}\label{sec:rq}

\subsection{Motivating Examples}
Bavota\etal\cite{DBLP:conf/msr/BavotaR16} conducted an empirical study on SATD using 159 repositories from Eclipse and Apache projects. They manually classified 366 SATDs into six categories: Code, Design, Document, Defect, Test, and Requirement. Code debt accounted for the largest proportion (30\%), followed by Defect debt and Requirement debt, each accounting for 20\%. Design, Document, and Test debt accounted for 12\%, 10\%, and 8\%, respectively. 
More importantly, most of the categories have subcategories except Test debt. For example, Code debt has two subcategories: ``low internal quality'' indicating poor code quality, and ``workaround'' referring to compromised code due to temporary implementations. 

In software projects, we have noticed that SATD comments in the test code do have different intentions: 
\begin{itemize}
\item {\bf Test failures}\\
\texttt{// TODO: Passes on macOS, fails on Linux and
Windows with AccessDeniedException.}
\item {\bf Need for specific tests}\\
\texttt{// TODO : more tests for datetimes with timezones and/or offsets}
\item {\bf Special implementation for testing purposes}\\
\texttt{// TODO: This is a hack, wc.login does not work with the form}
\end{itemize}

Some studies have also observed a very small number of instances relevant to testing when they studied SATD mainly in production code. Farias\etal\cite{DBLP:journals/infsof/FariasNKS20} proposed a pattern-based method for detecting SATD and conducted an empirical evaluation on three repositories: JEdit, Lucene, and ArgoUML. During the evaluation, they observed three types of Test debt: ``deficiencies in testing activities'' indicating defects in testing, ``tests to do'' indicating tests that should be conducted, and ``insufficient code coverage'' indicating low test coverage. However, these subcategories were derived from a small number of samples (\ie only 9 instances). 
Kashiwa\etal\cite{DBLP:journals/infsof/KashiwaNKKSSU22} also reported test-related SATD when investigating the impact of SATD in modern code reviews. They randomly sampled 375 SATD comments from OpenStack and Qt projects and classified them into six types: Scheduling, Work Dependency, Communication, Problem Report, Workaround, and Test. They observed 10 instances in the Test category and identified two subcategories: Necessity and Failure. The former refers to the cases where sufficient tests are missing for the method, while the latter indicates that tests fail at the location containing the SATD. 
Azuma\etal\cite{DBLP:journals/ese/AzumaMKK22} investigated SATD in Docker using Dockerfiles collected from the top 1250 images on Docker Hub. 
They manually classified 382 comments and found 50 SATD instances, which were further categorized into five types: Code Debt, Test Debt, Defect Debt, Design Debt, and Process Debt. Two subcategories were identified for Test debt: ``integrity check'' (lack of an integrity check on binary files or hash values used in a container) and ``improvement for test'' (asking for improvements in testing methods). According to their study, while there were only 9 cases of SATD related to testing, Test Debt was the second most common type of SATD after Code Debt, which had the most cases (26). 

While researchers have created several subcategories for test debt, they are normally derived from a small size of samples. 
Moreover, they are mainly a side product when studying SATD in whole software projects with no specific attention to testing.  
In this study, we collected 2,779 SATD instances present in the test code from 50 repositories to reveal the prevalence of Test SATD. We also manually classified 506 samples to disclose the intentions behind Test SATD. Studying SATD in test code can help us understand what test-related issues developers recognize and how they are handled. Moreover, we might identify new test smells with Test SATD. 

\subsection{Proposed Research Questions}

In this study, we aim to answer the following four research questions (RQs). 
Note that this study refers to SATD in production code as \texttt{Production SATD} and SATD in test code as \texttt{\ts}.

\smallskip
\noindent
{\bf \rqA} \\ \indent

A previous study by Bavota\etal\cite{DBLP:conf/msr/BavotaR16} has claimed that test-related SATD is the least frequent one by manually inspecting 366 SATD instances. However, the SATD instances were collected from only two ecosystems (Apache and Eclipse), which hinders the generalizability of the finding. 
Additionally, the original study was conducted in 2016, and there are currently more advanced SATD detection tools available, which might lead to different results. 
As the first step of this study, we aim to replicate the study of Bavota\etal to reveal the prevalence of SATD in test code, collected from a much more diverse repositories.

\smallskip
\noindent
{\bf \rqB} \\ \indent
Previous research on SATD has investigated the relationship between SATD and various software quality indicators, such as complexity metrics~\cite{DBLP:conf/icsm/PotdarS14}, coupling and readability~\cite{DBLP:conf/msr/BavotaR16}, code smells~\cite{DBLP:journals/tse/MaldonadoST17, DBLP:journals/sqj/RantalaML24}.
These studies suggest that SATD is not directly correlated with many aspects of code quality, although there is certain correlation between SATD and code smells. 
It is worth noting that these studies mainly target SATD in production code, and there are currently no studies examining the relationship between Test SATD and test quality. In RQ2, we aim to fill this gap.  

\smallskip
\noindent
{\bf \rqC} \\ \indent
Existing studies \cite{DBLP:conf/msr/BavotaR16, DBLP:journals/infsof/KashiwaNKKSSU22} have identified several test-related SATD. However, they are either too coarse-grained or derived from a small size of samples. 
To get a comprehensive picture of what types of SATD exist in the test code, in RQ3, we aim to identify the Test SATD types in a systematic manner by manually analyzing Test SATD instances and categorizing them based on the intentions of developers. 

\smallskip
\noindent
{\bf \rqD} \\ \indent
Various approaches~\cite{DBLP:journals/tse/MaldonadoST17, DBLP:journals/tr/ChenYFWC22, DBLP:conf/ease/SalaTF21} have been proposed for detecting and classifying SATD. 
However, these tools mainly focus on production code and no tool has been developed to categorize SATD in test code. In RQ4, we aim to fill this gap and develop a classifier to categorize Test SATD based on the categories defined in \rqC. 
We believe that the new model could streamline future analysis of Test SATD. 
\section{Methodology}
\label{sec:Method}
This section describes the data collection and analysis process in this study. 

\subsection{Data Collection}
\label{subsec:collect-repo}
\begin{figure*}[t]
    \centering
    \includegraphics[width=\linewidth]{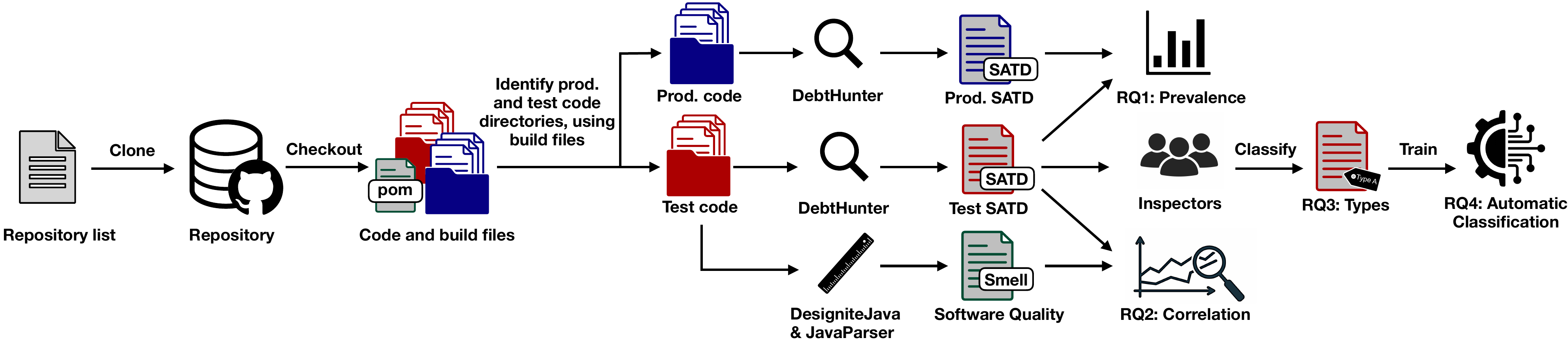}
    \caption{Overview of the data analysis}
    \label{fig:method}
\end{figure*}

Figure \ref{fig:method} shows the overview of our data collection process. To collect SATD instances from diverse software repositories, we select 50 projects: 20 projects from previous SATD studies~\cite{DBLP:journals/tse/MaldonadoST17,DBLP:journals/corr/abs-1910-13238} and 30 projects from testing-related studies~\cite{DBLP:conf/msr/BuiSF22, DBLP:conf/icsm/KashiwaS0BLKU21, DBLP:conf/scam/MartinsCRPM23, DBLP:journals/tse/SoaresRGAS23}. This is in line with our goal to study Test SATD as previous SATD studies mainly focus on production code. 
Table~\ref{tab:repositories} summarizes the repositories used in this study.

\begin{table}[p]
    \centering
    \footnotesize
    \caption{Statistics of the studied repositories}
    \label{tab:repositories}
    \begin{tabular}{rp{2.1cm}rrrrrrrrl}
        \toprule
        \# & Repository & P.LOC & T.LOC & Files & SATDs & Stars & Commits & Forks & Dev. & Ref \\
        \midrule
        1 & Ant & 114,501 & 32,279 & 1,327 & 243 & 442 & 14,963 & 445 & 80 & \cite{DBLP:journals/tse/MaldonadoST17} \\
        2 & ArgoUML & 160,769 & 16,705 & 1,908 & 1,828 & 262 & 17,797 & 102 & 5 & \cite{DBLP:journals/tse/MaldonadoST17} \\
        3 & Columba & 66,394 & 1,526 & 993 & 138 & - & - & - & - & \cite{DBLP:journals/tse/MaldonadoST17} \\
        4 & EMF & 606,645 & 105,084 & 3,140 & 141 & 20 & 10,341 & 24 & 36 & \cite{DBLP:journals/tse/MaldonadoST17} \\
        5 & Hibernate & 163,448 & 166,085 & 4,705 & 812 & - & - & - & - & \cite{DBLP:journals/tse/MaldonadoST17} \\
        6 & Jedit & 121,532 & 3,502 & 617 & 164 & - & - & - & - & \cite{DBLP:journals/tse/MaldonadoST17} \\
        7 & JFreeChart & 97,460 & 41,664 & 1,017 & 93 & 1,307 & 3,894 & 565 & 30 & \cite{DBLP:journals/tse/MaldonadoST17} \\
        8 & Jmeter & 120,620 & 28,143 & 1,402 & 272 & 8,921 & 18,254 & 2,193 & 80 & \cite{DBLP:journals/tse/MaldonadoST17} \\
        9 & Jruby & 265,506 & 10,427 & 1,793 & 1,016 & 3,835 & 53,597 & 928 & 362 & \cite{DBLP:journals/tse/MaldonadoST17} \\
        10 & Squirrel& 186,681 & 28,553 & 2,325 & 253 & 75 & 7,766 & 19 & 20 & \cite{DBLP:journals/tse/MaldonadoST17} \\
        11 & Dubbo& 154,678 & 108,908 & 3,581 & 168 & 41,184 & 7,535 & 26,527 & 401 & \cite{DBLP:journals/corr/abs-1910-13238} \\
        12 & Gradle & 494,771 & 17,712 & 10,392 & 698 & 17,886 & 114,478 & 4,984 & 345 & \cite{DBLP:journals/corr/abs-1910-13238} \\
        13 & Groovy & 191,447 & 12,262 & 1,793 & 462 & 5,336 & 20,871 & 1,911 & 359 & \cite{DBLP:journals/corr/abs-1910-13238} \\
        14 & Hive & 867,928 & 178,970 & 5,086 & 1,354 & 5,747 & 10,189 & 4,757 & 257 & \cite{DBLP:journals/corr/abs-1910-13238} \\
        15 & Maven & 55,245 & 19,455 & 1,012 & 173 & 4,716 & 10,966 & 2,764 & 209 & \cite{DBLP:journals/corr/abs-1910-13238} \\
        16 & Poi & 281,995 & 142,878 & 3,710 & 818 & 2,062 & 12,828 & 792 & 17 & \cite{DBLP:journals/corr/abs-1910-13238} \\
        17 & Spring Framework & 383,270 & 391,933 & 8,310 & 283 & 58,540 & 29,550 & 38,600 & 355 & \cite{DBLP:journals/corr/abs-1910-13238} \\
        18 & Storm & 249,986 & 40,390 & 2,145 & 103 & 6,644 & 10,798 & 4,064 & 281 & \cite{DBLP:journals/corr/abs-1910-13238} \\
        19 & Tomcat & 268,488 & 92,832 & 2,616 & 814 & 7,891 & 25,797 & 5,211 & 133 & \cite{DBLP:journals/corr/abs-1910-13238} \\
        20 & Zookeeper & 61,382 & 59,789 & 925 & 109 & 12,549 & 2,534 & 7,295 & 238 & \cite{DBLP:journals/corr/abs-1910-13238} \\
        21 & Commons IO & 17,890 & 33,866 & 513 & 87 & 1,036 & 4,849 & 695 & 115 & \cite{DBLP:conf/icsm/KashiwaS0BLKU21} \\
        22 & Spring & 2,079 & 4,240 & 123 & 5 & 2,886 & 2,165 & 2,628 & 49 & \cite{DBLP:conf/icsm/KashiwaS0BLKU21} \\
        23 & Joda-Beans & 17,353 & 34,625 & 328 & 2 & 146 & 891 & 40 & 11 & \cite{DBLP:conf/icsm/KashiwaS0BLKU21} \\
        24 & Jsoup & 15,306 & 14,745 & 153 & 58 & 11,222 & 1,969 & 2,248 & 103 & \cite{DBLP:conf/icsm/KashiwaS0BLKU21} \\
        25 & Spark & 6,222 & 5,048 & 184 & 5 & 9,660 & 1,041 & 1,569 & 123 & \cite{DBLP:conf/icsm/KashiwaS0BLKU21} \\
        26 & LittleProxy & 4,180 & 4,665 & 88 & 3 & 2,095 & 998 & 782 & 27 & \cite{DBLP:conf/icsm/KashiwaS0BLKU21} \\
        27 & RxJava JDBC & 4,611 & 3,340 & 79 & 7 & 803 & 926 & 115 & 13 & \cite{DBLP:conf/icsm/KashiwaS0BLKU21} \\
        28 & Spoon & 83,543 & 83,893 & 2,407 & 148 & 1,840 & 4,651 & 361 & 134 & \cite{DBLP:conf/icsm/KashiwaS0BLKU21} \\
        29 & Accumulo & 441,912 & 44,357 & 2,179 & 110 & 1,104 & 9,001 & 463 & 158 & \cite{DBLP:conf/scam/MartinsCRPM23} \\
        30 & BookKeeper & 157,229 & 121,242 & 2,351 & 62 & 1,955 & 3,322 & 960 & 204 & \cite{DBLP:conf/scam/MartinsCRPM23} \\
        31 & Camel & 1,041,075 & 653,891 & 21,464 & 1,999 & 5,906 & 70,277 & 5,048 & 331 & \cite{DBLP:conf/scam/MartinsCRPM23} \\
        32 & Cassandra & 314,778 & 273,241 & 3,749 & 496 & 9,292 & 28,567 & 3,723 & 272& \cite{DBLP:conf/scam/MartinsCRPM23} \\
        33 & CXF & 389,365 & 308,696 & 7,513 & 443 & 885 & 18,004 & 1,442 & 238 & \cite{DBLP:conf/scam/MartinsCRPM23} \\
        34 & Flink & 838,513 & 748,068 & 13,521 & 696 & 25,086 & 34,899 & 13,717 & 285 & \cite{DBLP:conf/scam/MartinsCRPM23} \\
        35 & Hadoop & 997,669 & 902,657 & 11,871 & 1,022 & 15,190 & 27,121 & 9,076 & 303 & \cite{DBLP:conf/scam/MartinsCRPM23} \\
        36 & Kafka & 271,653 & 356,138 & 4,245 & 327 & 30,571 & 11,728 & 14,524 & 347 & \cite{DBLP:conf/scam/MartinsCRPM23} \\
        37 & Karaf & 105,567 & 23,311 & 1,603 & 135 & 691 & 9,471 & 664 & 163 & \cite{DBLP:conf/scam/MartinsCRPM23} \\
        38 & Wicket & 142,446 & 77,054 & 3,306 & 141 & 768 & 21,683 & 393 & 116 & \cite{DBLP:conf/scam/MartinsCRPM23} \\
        39 & Struts & 123,978 & 87,049 & 2,589 & 179 & 1,318 & 7,160 & 824 & 77 & \cite{DBLP:conf/msr/BuiSF22} \\
        40 & Compress & 43,544 & 29,119 & 623 & 170 & 370 & 5,121 & 295 & 95 & \cite{DBLP:conf/msr/BuiSF22} \\
        41 & Jenkins & 113,619 & 75,714 & 1,782 & 509 & 24,213 & 34,958 & 9,093 & 296 & \cite{DBLP:conf/msr/BuiSF22} \\
        42 & Spring Security & 103,939 & 167,283 & 3,145 & 64 & 9,224 & 15,713 & 6,092 & 388 & \cite{DBLP:conf/msr/BuiSF22} \\
        43 & FileUpload & 2,221 & 4,205 & 92 & 13 & 248 & 1,657 & 187 & 49 & \cite{DBLP:conf/msr/BuiSF22} \\
        44 & Imaging & 30,710 & 11,347 & 623 & 73 & 461 & 2,486 & 197 & 45 & \cite{DBLP:conf/msr/BuiSF22} \\
        45 & Sling & 7,246 & 5,750 & 179 & 8 & 15 & 687 & 22 & 21 & \cite{DBLP:conf/msr/BuiSF22} \\
        46 & UAA & 62,053 & 124,466 & 1,646 & 24 & 1,625 & 10,952 & 834 & 158 & \cite{DBLP:conf/msr/BuiSF22} \\
        47 & Jackson& 6,283 & 10,702 & 195 & 9 & 600 & 1,560 & 230 & 41 & \cite{DBLP:conf/msr/BuiSF22} \\
        48 & Prime JWT & 3,759 & 2,809 & 101 & 4 & 187 & 326 & 43 & 12 & \cite{DBLP:conf/msr/BuiSF22} \\
        49 & OpenRefine & 57,094 & 28,418 & 1,026 & 213 & 11,446 & 7,995 & 2,080 & 353 & \cite{DBLP:conf/msr/BuiSF22} \\
        50 & Quarkus & 540,637 & 449,513 & 16,729 & 812 & 14,764 & 45,217 & 2,900 & 391 & \cite{DBLP:journals/tse/SoaresRGAS23} \\
        \bottomrule
    \end{tabular}
\end{table}

For each of these repositories, we clone and checkout the latest revision of the master/main branch. However, for some repositories, we collected specific snapshot versions instead. We then identify SATD in the production code and the test code.
Since all 50 repositories in our study are Java projects using either Maven or Gradle, we identified production and test directories by analyzing their build configuration files. We located each repository's \texttt{pom.xml} file (Maven projects) or \texttt{build.gradle} file (Gradle projects) and parsed it to extract directory paths. For Maven projects, we extracted paths from the \texttt{sourceDirectory} and \texttt{testSourceDirectory} elements; for Gradle projects, we identified paths from the \texttt{main} and \texttt{test} properties. When these paths were not explicitly defined, we used standard Java conventions: \texttt{src/main/java} for production code and \texttt{src/test/java} for test code.\footnote{\url{https://maven.apache.org/guides/introduction/introduction-to-the-standard-directory-layout.html}}%
\textsuperscript{,}%
\footnote{\url{https://docs.gradle.org/current/userguide/java_plugin.html\#sec:java_project_layout}}
In the end, we identify 100,364 files in the production directories and 60,515 files in the test directories. 
From these files, we identify SATDs in the production code and test code.
We consider SATD instances found in the production code as \texttt{Production SATDs} and those found in the test code as \texttt{Test SATDs}.
For SATD detection, we employ a state-of-the-art SATD detection tool, \verb|DebtHunter|~\cite{DBLP:conf/ease/SalaTF21}. \verb|DebtHunter| is an ML-based tool that uses the Sequential Minimal Optimization (SMO) algorithm~\cite{DBLP:journals/jair/ChawlaBHK02}. 
A previous study~\cite{DBLP:conf/ease/SalaTF21} has demonstrated the high performance of \verb|DebtHunter|, with a precision of 0.972, recall of 0.967, and F1-score of 0.965, outperforming other SATD detection tools. Additionally, since DebtHunter has been trained with multiple labels for classifying the types of SATD, we also use this tool to identify SATD types. Note that we modified the source code of \verb|DebtHunter| so that it can detect Test SATD because the original version excludes the test code.

\subsection{Data Analysis}

\subsubsection{RQ1 (Prevalence of \ts)}
\begin{figure}
    \centering
    \includegraphics[width=0.9\linewidth]{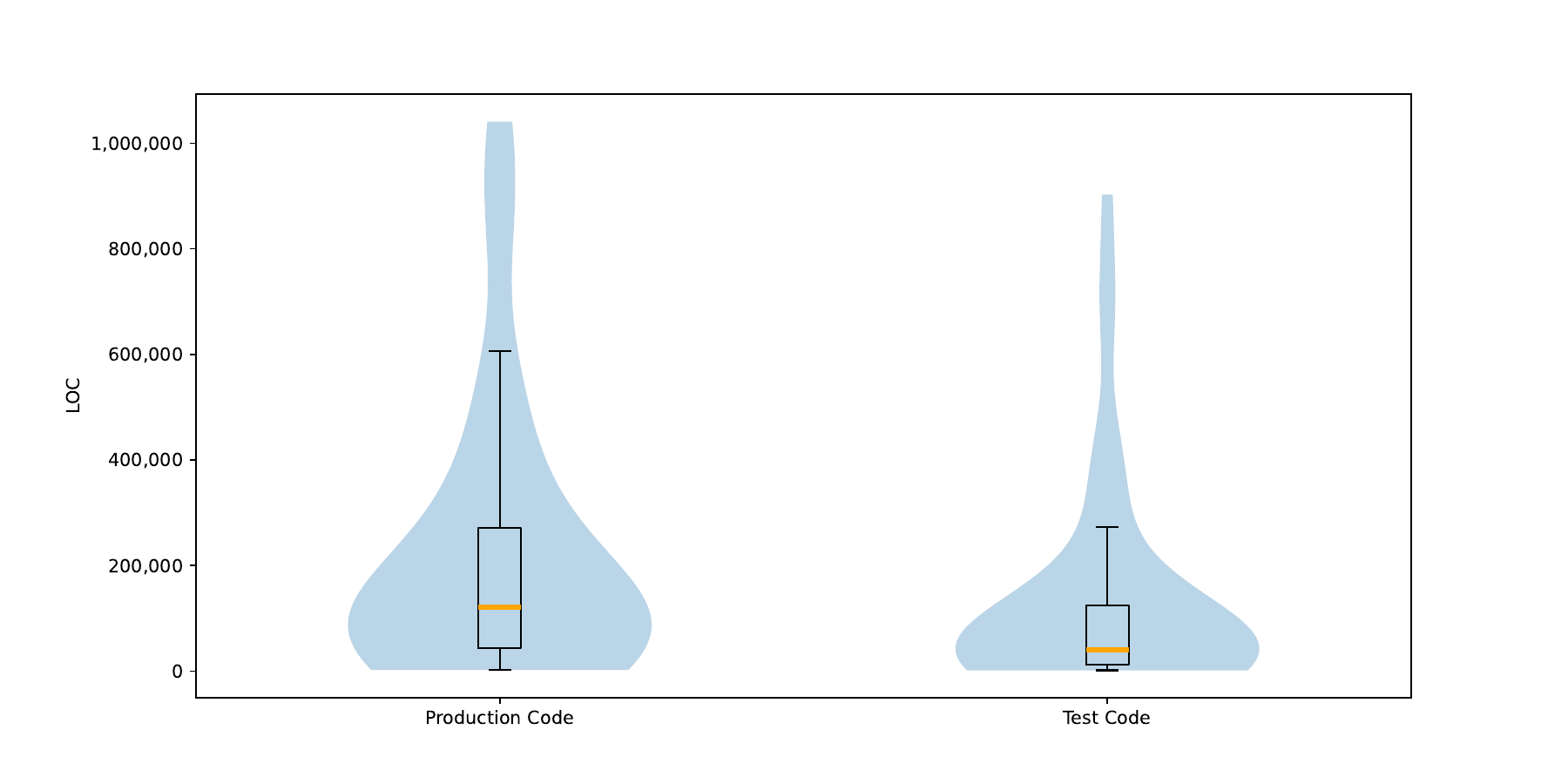}
    \caption{Distributions of LOC in the production and test Code}
    \label{fig:violin_loc}
\end{figure}

To understand how prevalent SATD is in the test code, we count the number of Test SATD and compare it with that of Production SATD. In software projects, the size (\ie lines of code) of the production code and test code differ significantly.
Figure \ref{fig:violin_loc} shows the distribution of LOC of the production code and test code. The median numbers of LOC in the production code and test code are 121,076 and 37,507.5, respectively.
To ensure a fair comparison, we measure the number of SATD instances per 10,000 lines of code. Finally, we apply the Wilcoxon signed-rank test~\cite{Wilcoxon1945}, which is a non-parametric test to compare paired data and confirm a statistically significant difference ($\alpha < 0.01$). We also measure the effect size using Cliff's Delta ($d$)~\cite{grissom2005effect}. 
We follow the well-established guideline~\cite{grissom2005effect} to interpret the effect size: negligible
for $|d| < 0.10$, small for $0.10 \le |d| < 0.33$, medium for
$0.33 \le |d| < 0.474$, and large for $|d| \ge 0.474$. 


\subsubsection{RQ2 (Relationship with Test Quality) }
To understand how Test SATD is correlated with the test quality, we use several proxies to represent the test quality, including general code quality metrics (\codesmells, lines of code (LOC), complexity, and readability) and test-specific metrics (\testsmells, assertions, and annotations) at the method level. 
The LOC, the number of assertions, and annotations are counted after parsing the source code with \verb|JavaParser|.\footnote{\url{https://github.com/javaparser/javaparser}}

The test smells are detected with \verb|tsDetect|~\cite{DBLP:conf/sigsoft/PerumaANM0P20},\footnote{\url{https://github.com/bin-lin/TestSmellDetector}} a state-of-the-art tool which can detect 16 of 19 test smells proposed by van Deursen\etal~\cite{DBLP:article/deursen/2001} at the method level. Their evaluation shows an average precision and recall exceeding 95\% for detecting each type of smell. To detect code smells, we use \verb|Designite Java|,\footnote{\url{https://github.com/tushartushar/DesigniteJava}} which can detect 10 of 11 \codesmells proposed by Sharma\etal~\cite{DBLP:conf/esem/SharmaFS17}.
To ensure the reliability of these automated detectors on our specific dataset, we conducted an experimental manual validation. We randomly sampled 100 instances identified as test smells by \verb|tsDetect| and 100 instances identified as code smells by \verb|Designite Java|. Two of the authors then independently inspected each instance to verify if it was a true positive. Any disagreements were resolved through
discussion with a third inspector to reach a consensus. Our validation confirmed a high precision for both tools: 90\% for the test smell detection and 94\% for the code smell detection. These results to a high extent guarantee the validity of the quality metrics used in our analysis.

We also use \verb|Designite Java| to calculate the cyclomatic complexity of each method, which is represented as an integer with a minimum value of 1, where a higher value indicates greater complexity. 
For readability, we use the tool proposed by Scalabrino\etal\cite{DBLP:journals/smr/ScalabrinoLOP18}. This tool utilizes a comprehensive model that integrates textual features, such as identifiers and comments, with structural code features. The readability score is evaluated on a scale from 0 to 1, where a higher value indicates better readability. 
Note that code comments increase the readability score, and SATD comments may increase the score as well. To examine whether SATDs are included in the methods with low readability independent of SATD comment influence, we removed SATD comments in the test methods before applying the tool.

Next, these data are matched with the same method to investigate the relationship between various software qualities and Test SATD. For test smells and code smells, Chi-square test~\cite{tallarida1987chi} is conducted to examine the relationship between the presence or absence of each smell. The null hypothesis is set as ``The presence or absence of Test SATD and the presence or absence of each smell are independent,'' and tested at the significance level of $\alpha = 0.05$. The presence or absence of Test SATD was determined at the method level, based on whether SATD exists in the documentation comments written for the method or in the comments contained within the method. Furthermore, to investigate whether there is a difference in the number of each smell based on the presence or absence of Test SATD, a non-parametric test, the Mann-Whitney U test~\cite{mann1947test}, is used to calculate the p-value. The null hypothesis is ``There is no difference in the number of smells based on the presence or absence of Test SATD,'' and the test is conducted with a significance level of $\alpha = 0.05$. In addition, we used the U-score obtained during the computation of the Mann-Whitney U test to obtain a Z-score and calculate the effect size. The effect size r \cite{mangiafico2016summary} is determined using the formula $r = |z|/\sqrt{Sample Size}$ and ranges between 0 and 1. Generally, an effect size $r$ between 0.1 and 0.3 indicates a small effect, values greater than 0.3 and up to 0.5 indicate a medium effect, and values greater than 0.5 indicate a large effect~\cite{coolican2017research}. 
Similarly, we measure LOC, the number of assertions, annotations, cyclomatic complexity, and readability, when the test method does/does not contain Test SATDs. And then we examine the statistically significant difference between these two groups using the Mann-Whitney U test.

\subsubsection{RQ3 (Purposes of \ts)}
\label{sec:rq3-method}
To understand developers' intentions behind Test SATD, we perform a manual inspection to categorize Test SATD. To facilitate the classification, we apply several filters to exclude comments that do not contain meaningful content from the collected Test SATD in the data collection process. Specifically, we remove comments that contain only tags like \verb|TODO| or \verb|FIXME| without any other text. This filtering eliminates 651 SATD instances from the original 2,779 instances. Additionally, we remove comments that are automatically inserted by integrated development environments (IDEs) such as \verb|TODO Auto-generated method stub|. This step removes 36 SATD instances, leading to 2,092 remaining.

Next, we randomly select \SampleSize SATD comments, which represent a statistically significant sample with a 99\% confidence level and a 5\% confidence interval. 
For the classification, we adopted an inductive coding approach, allowing categories to emerge directly from the data rather than starting with pre-existing taxonomies from studies on production code or non-self-admitted technical debt. This methodological decision was made to avoid categorization bias, as pre-existing categories developed from production code might overlook patterns and issues unique to the test code context. By developing the taxonomy from the ground up, we aimed to ensure that it precisely represents how developers document debt in testing contexts and to enable the discovery of test-specific insights.

The classification itself was conducted using a multi-stage, iterative approach based on card sorting~\cite{spencer2009card}, consisting of five iterations with approximately 100 instances each. In the first iteration, three authors independently assigned free-form labels to the initial 100 SATD instances to capture their intent, allowing for variations in wording for similar concepts. Afterward, the three authors discussed these labels. By resolving disagreements and grouping semantically similar labels, they created an initial set of categories. In the subsequent four iterations, two authors independently classified the remaining samples (in batches of approximately 100) using this initial category set. A third author solved any disagreements that arose through the discussion with the others. If an instance did not fit any existing category, a new category was proposed and added to the set after a discussion among the three authors. The three annotators involved in this process have 5-15 years of programming experience.

\subsubsection{RQ4 (Automatic classification)}
To examine the extent to which we can accurately classify the SATD types, in RQ4, we construct a classifier following the approach by Sabbah\etal~\cite{sabbah2023self} that classifies production SATD types using natural language processing word embeddings. We utilize the comment texts and labels assigned in RQ3
for the training and testing process. 

Prior to training, we pre-process the comments by removing punctuation marks and HTML tags. We then tokenize the comments, remove stop words, and perform stemming. The tokens are converted into numerical vectors using Term Frequency-Inverse Document Frequency (TF-IDF)~\cite{DBLP:journals/jd/Jones04}. 
We also employ various machine-learning algorithms used in the previous studies~\cite{DBLP:journals/tr/ChenYFWC22, DBLP:journals/ese/HuangSXLL18, DBLP:journals/access/FlisarP19}, including Support Vector Machines (SVM)~\cite{vapnik1963}, Naive Bayes (NB)~\cite{kononenko1990comparison}, Random Forest~\cite{DBLP:journals/ml/Breiman01} and eXtreme Gradient Boosting (XGBoost)~\cite{DBLP:conf/kdd/ChenG16}. 

Additionally, we developed language model-based classifiers due to their high classification accuracy demonstrated in previous studies~\cite{sabbah2023self}. Specifically, we employed BERT~\cite{DBLP:conf/naacl/DevlinCLT19} and CodeBERT~\cite{DBLP:conf/emnlp/FengGTDFGS0LJZ20}, a pre-trained model capable of understanding the context of source code and natural language.
In our study, we replace the output layer of the model with a linear layer to enable the model to classify SATD comments into five sub-categories of \ts. We employ GELU for the activation function and AdamW for the optimizer. The maximum token length is set to 128, the batch size to 16, and the maximum number of epochs to 20. To prevent overfitting~\cite{DBLP:journals/jcisd/TetkoLL95} during training, we employ early stopping~\cite{Yuan2007}, which halts training when the validation loss does not improve for five consecutive epochs.

Furthermore, we measure the performance of Large Language Models (LLMs) for Test SATD classification. This was inspired by a recent study by Li\etal\cite{DBLP:journals/spe/LiLLYLK25}, which investigates the capabilities of generative models such as ChatGPT for SATD detection. We constructed classifiers using two models via the OpenAI API: \texttt{GPT-3.5-turbo} and \texttt{GPT-4.1}. Unlike the fine-tuning approach used for CodeBERT, we employed a few-shot prompting strategy. For each SATD instance to be classified, the prompt included a description of our five SATD categories along with typical examples of each to guide the model's reasoning. The performance of these LLM-based classifiers was evaluated on the same dataset using the same metrics to ensure a direct comparison with our other models.

To assess model performance, we employed 10-fold cross-validation~\cite{DBLP:journals/neco/Dietterich98}, a widely adopted technique for robust evaluation. The dataset was randomly partitioned into ten equally sized folds. In each iteration, nine folds were used for training and the remaining fold for testing, ensuring that each fold served as the test set exactly once. For LLM-based classifiers (GPT-3.5-turbo and GPT-4.1), which do not require training data, we applied the same 10-fold partitioning scheme but used only the test folds for evaluation.
Across all models, we measure performance using three classic metrics: Precision, Recall, and F1-score.
The F1 score is the harmonic mean of Precision and Recall. Since there is a trade-off between Precision and Recall, the F1 score evaluates the balance between Precision and Recall. In other words, it assesses whether the increase in Precision (or Recall) outweighs the decrease in Recall (or Precision).


\section{Results}
\label{sec:Result}


\begin{figure}[t]
    \centering
    \includegraphics[width=0.85\linewidth]{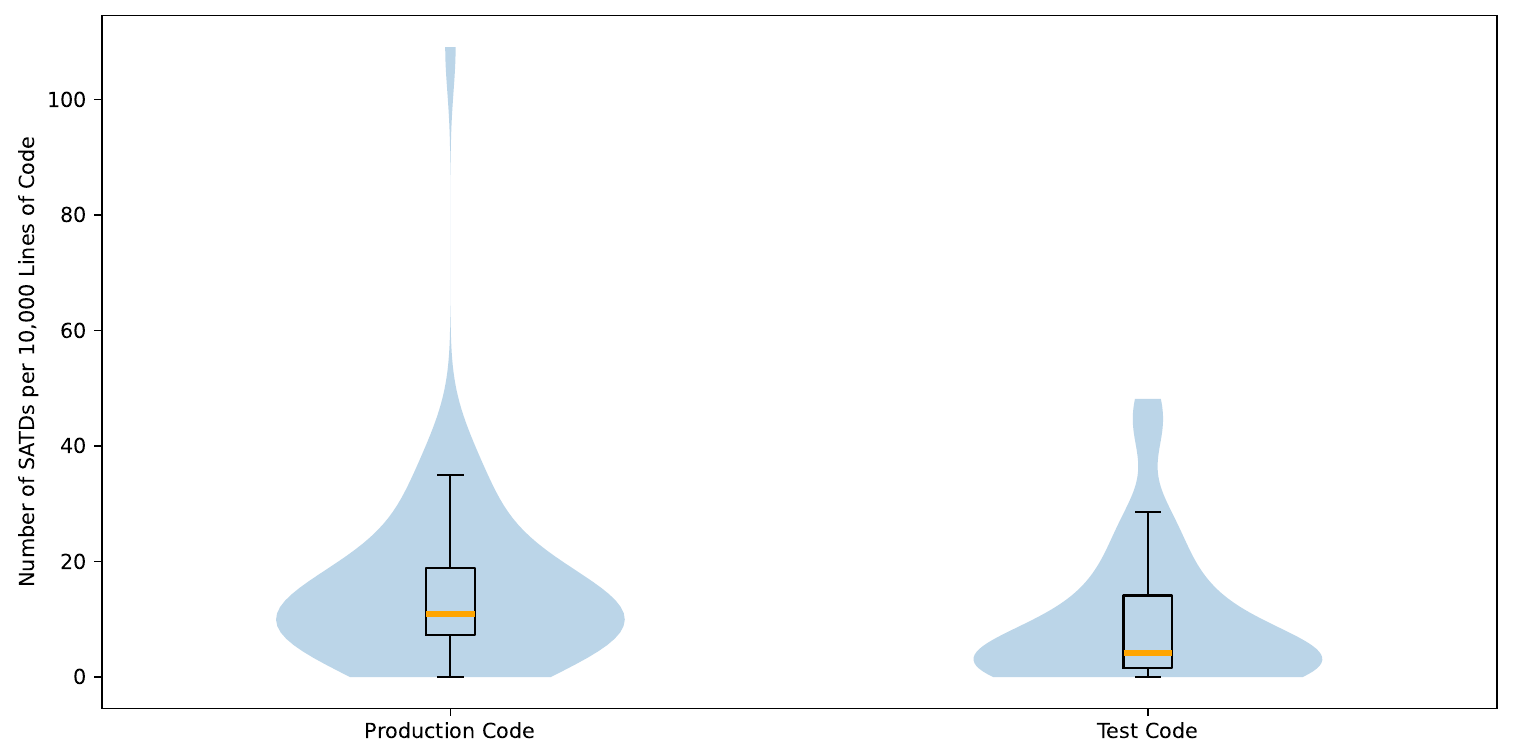}
    \caption{Violin Plot of \# SATD/10kLOC in Production and Test Code}
    \label{fig:violin_satd}
\end{figure}
\subsection{\rqA}
By applying the SATD detection tool \texttt{DebtHunter}, we obtained a total of 14,987 and 2,779 SATD instances in the production and test code, respectively. SATD in test code accounts for 15.6\% of all detected SATD, which differs from the result reported in Bavota\etal~\cite{DBLP:conf/msr/BavotaR16}. 
Their finding indicates that test-related SATD accounts for only 8\%, significantly lower than our results. This discrepancy may be caused by three factors. First, we collected SATD instances from different repositories, and half of them are used in testing-related studies, which might indicate that the repositories are well-tested. Second, as the repositories in the previous study were collected in 2015 (those in ours collected in 2024), writing tests have become more common recently~\cite{DBLP:journals/access/GurcanDCRS22, DBLP:conf/mipro/HynninenKKT18}.
Third, the previous studies manually analyzed whether the SATD is directly related to testing activities, while we count SATD instances in test code instead. For example, if SATD is related to software implementation, it will not fall into the test category even if it is located in the test code. 

As the lines of code in the production code and test code differ significantly, we also compare the normalized number of SATD instances. Figure \ref{fig:violin_satd} depicts the distribution of the numbers of SATD per 10,000 lines of code for each repository. In these violin plots, the thickness of the outer layer represents the probability density of the plotted values. In the center of each violin plot, the yellow line shows the median while the box bar represents the interquartile range.
When looking at the quartiles, the number of SATD instances per 10k LOC range from \ProdOneQuartile to \ProdThreeQuartile in the production code, while from \TestOneQuartile to \TestThreeQuartile in the test code. The average and median values for Production SATD are also much higher than that for Test SATD (average: \ProdAve vs. \TestAve, median: \ProdMedian vs. \TestMedian). 
In terms of median, the normalized number of Test SATD is \PerDiff smaller than that of the Production SATDs. A statistically significant difference is also observed with the Wilcoxon signed-rank test (p-value: 0.00006). In addition, the effect size (Cliff's Delta) was found to be \EffectSize, indicating a medium difference.
Table \ref{tab:statistical_summary} presents the detailed results of the statistical test.

\smallskip
\begin{tcolorbox}
\textbf{RQ1. While there are fewer Test SATD instances than Production SATD (\ie \ProdMedian and \TestMedian, respectively), the number is still non-negligible in the studied repositories.}
\end{tcolorbox}

\begin{table*}[h]
\centering
\caption{Summary of Key Statistical Test Results}
\label{tab:statistical_summary}
\begin{tabular}{llccc}
\toprule
\textbf{} & \textbf{Metrics} & \textbf{Statistical Test} & \textbf{p-value} & \textbf{Effect Size} \\
\midrule
\textbf{RQ1} & SATD Density & Wilcoxon signed-rank & $< 0.001$ & Medium {\normalsize (d = 0.365)} \\
\midrule
\multirow{7}{*}{\textbf{RQ2}} & LOC & Mann-Whitney U & $< 0.001$ & Negligible {\normalsize (r = 0.009)} \\
& Assertions & Mann-Whitney U & $0.48$ &  Negligible {\normalsize (r = 0.009)} \\
& Annotations & Mann-Whitney U & $< 0.001$ & Negligible {\normalsize (r = 0.006)} \\
& Cyclomatic Comp.  & Mann-Whitney U & $< 0.001$ & Small {\normalsize (r = 0.014)} \\
& Readability  & Mann-Whitney U & $< 0.001$ & Negligible {\normalsize (r = 0.006)} \\
& Test Smells & Mann-Whitney U & $0.08$ & Negligible {\normalsize (r = 0.002)} \\
& Code Smells  & Mann-Whitney U & $< 0.001$ & Small {\normalsize (r = 0.013)} \\
\bottomrule
\end{tabular}
\end{table*}
\subsection{\rqB}
We calculated test quality metrics for test methods with and without Test SATD. Figure \ref{fig:distributions} depicts the distributions of each metric based on the presence or absence of Test SATDs.
Table \ref{tab:statistical_summary} presents the detailed results of this test, including the p-value and effect size. 

For lines of code (LOC) (Figure \ref{fig:loc}), test methods with SATD had a median LOC of 8, which is slightly higher than those without SATD (median LOC: 7). We applied the Mann-Whitney U test to these two groups and observed a statistically significant difference ($p = 2.32 \times 10^{-10}, r = 0.0094$). 
Note that the calculation of LOC excludes comments, indicating that test methods with SATD are simply larger.

Regarding the number of assertions (Figure \ref{fig:assertions}), the medians were both 0, and no statistically significant difference was observed ($p = 0.48, r = 0.00092$). However, for the number of annotations (Figure \ref{fig:annotations}), the medians were both 1, but on average, test methods with SATD had 0.79 annotations, whereas those without SATD had 0.74 annotations. The Mann-Whitney U test identified a statistically significant difference ($p = 2.40 \times 10^{-6}, r = 0.0059$).

Regarding readability (Figure \ref{fig:readability}), we found a counter-intuitive result: the median for test methods with SATD was 0.60, which is slightly higher than those without SATD (\ie 0.59). A statistically significant difference was observed with the Mann-Whitney U test ($p = 4.02 \times 10^{-5}$, $r = 0.006$).    

For cyclomatic complexity (Figure \ref{fig:cc}), while the median was 1 for both groups, the mean for test methods with SATD was 1.72, compared to 1.29 for methods without SATD. The Mann-Whitney U test confirmed a statistically significant difference ($p = 4.95 \times 10^{-48}, r = 0.014$), clarifying that methods with Test SATD tend to be structurally more complex.

For the number of test smells (Figure \ref{fig:testsmells}), the medians were both 0, and no statistically significant difference was observed by the Mann-Whitney U test ($p = 0.08, r = 0.0022$). Additionally, we conducted Chi-square test to examine the relationship between the presence of Test SATD and test smells in methods. The observed frequencies of Test SATD and test smells are summarized in Table \ref{table:testsmell-crosstab}. The chi-square test yielded $\chi^2(1, N = 447,040) = 1.19, p = 0.28$, indicating no statistically significant association between the presence of Test SATD and the presence of test smells.
\begin{table}[ht]
\caption{Confusion matrix of test methods with/without Test SATDs and \textbf{Test} Smells}
\label{table:testsmell-crosstab}
\begin{tabular}{rrr}
\toprule
             & With Test Smell & Without Test Smell \\ \midrule
With SATD    & 854              & 1,761               \\
Without SATD & 149,721          & 294,704           \\ \bottomrule
\end{tabular}
\end{table}

On the other hand, for the number of code smells, although the medians were both 0, the Mann-Whitney U test showed a p-value of $1.36 \times 10^{-24}$, indicating a statistically significant difference with a negligible effect ($r = 0.013$). This confirms that methods with Test SATD tend to have a statistically higher number of code smells, but the difference is not large. We also performed the Chi-square test to investigate the relationship between the presence of Test SATD and code smells (the metrics are summarized in Table~\ref{table:codesmell-crosstab}). The chi-square test yielded $\chi^2(1, N = 447,040) = 49.3, p = 2.25 \times 10^{-12}$, indicating that there is a statistically significant difference.
This contrast between the findings for test smells and code smells is one of the key insights of our study. Our results show that while there is a statistically significant correlation between the presence of Test SATD and the number of code smells, we found no such correlation with the number of test smells. This implies that Test SATDs are more likely to happen in test methods that have code quality issues rather than test-specific design problems. In other words, Test SATD and test smells represent different facets of quality issues in test code and can occur independently.

\begin{table}[h]
\caption{Confusion matrix of test methods with/without Test SATDs and \textbf{Code} Smells}
\label{table:codesmell-crosstab}
\begin{tabular}{rrr}
\toprule
             & With Code Smell & Without Code Smell \\ \midrule
With Test SATD    & 1,041             & 1,574               \\
Without Test SATD & 147,996           & 296,429           \\ \bottomrule
\end{tabular}
\end{table}

\begin{figure}[bp]
    \centering

    \begin{subfigure}{0.45\textwidth}
        \centering
        \includegraphics[width=\linewidth]{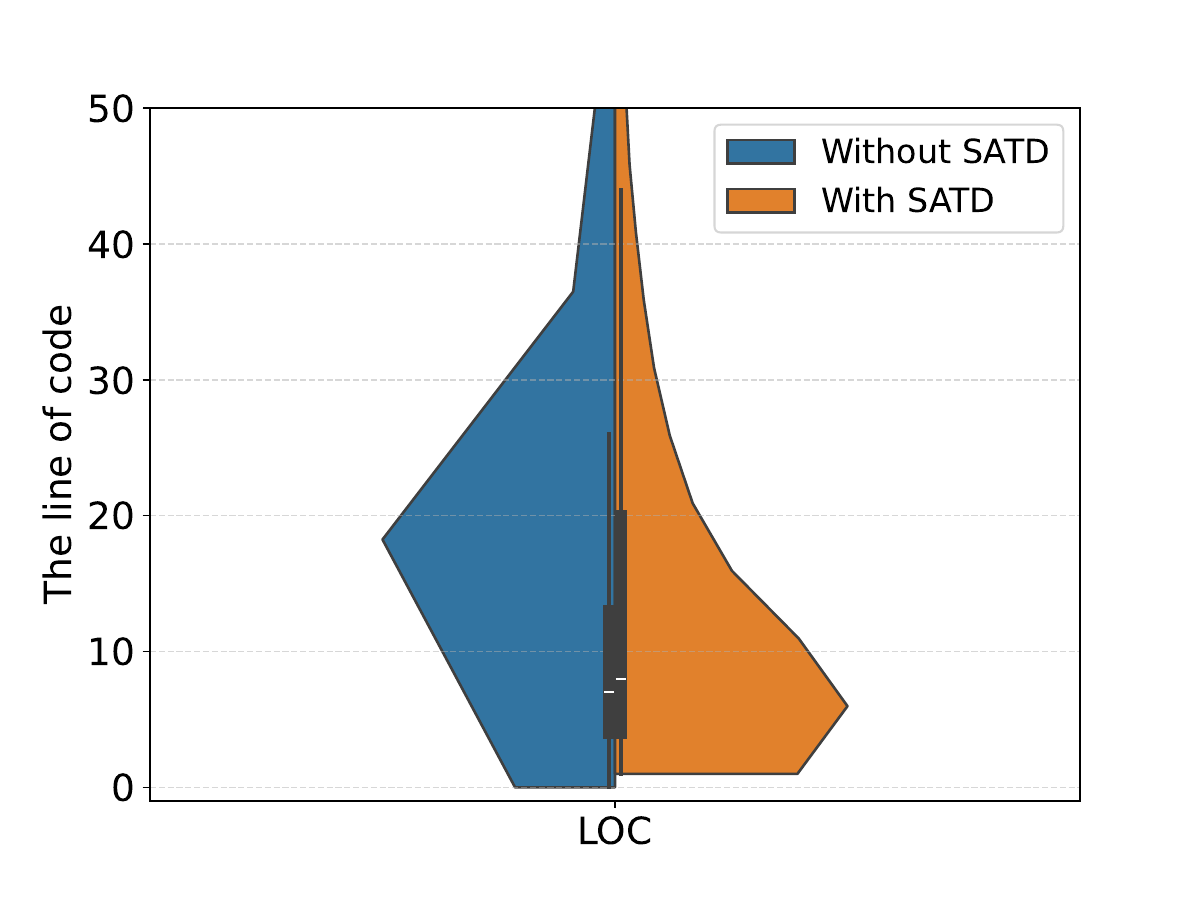}
        \caption{LOC}
        \label{fig:loc}
    \end{subfigure}
    \begin{subfigure}{0.45\textwidth}
        \centering
        \includegraphics[width=\linewidth]{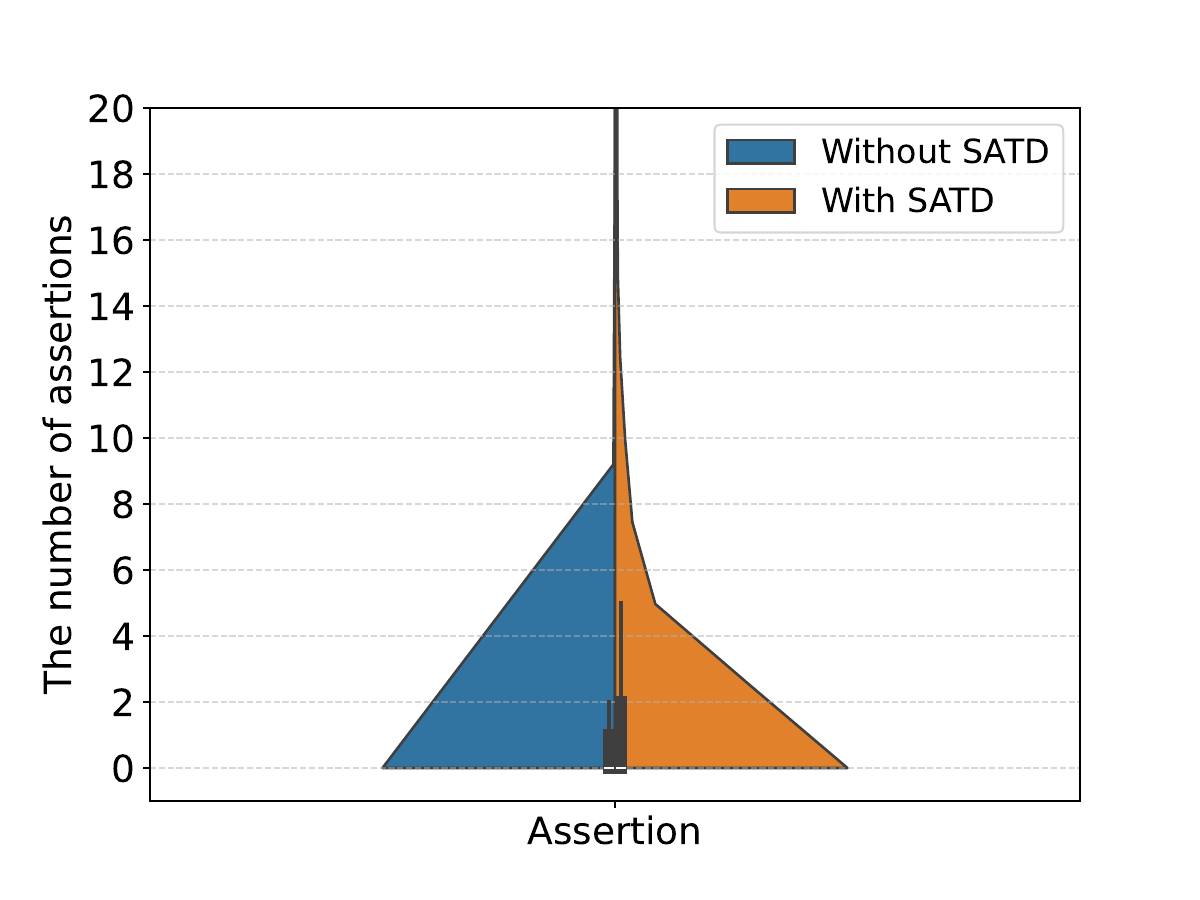}
        \caption{Assertions}
        \label{fig:assertions}
    \end{subfigure}
    
    \vspace{0.5cm}
    
    \begin{subfigure}{0.45\textwidth}
        \centering
        \includegraphics[width=\linewidth]{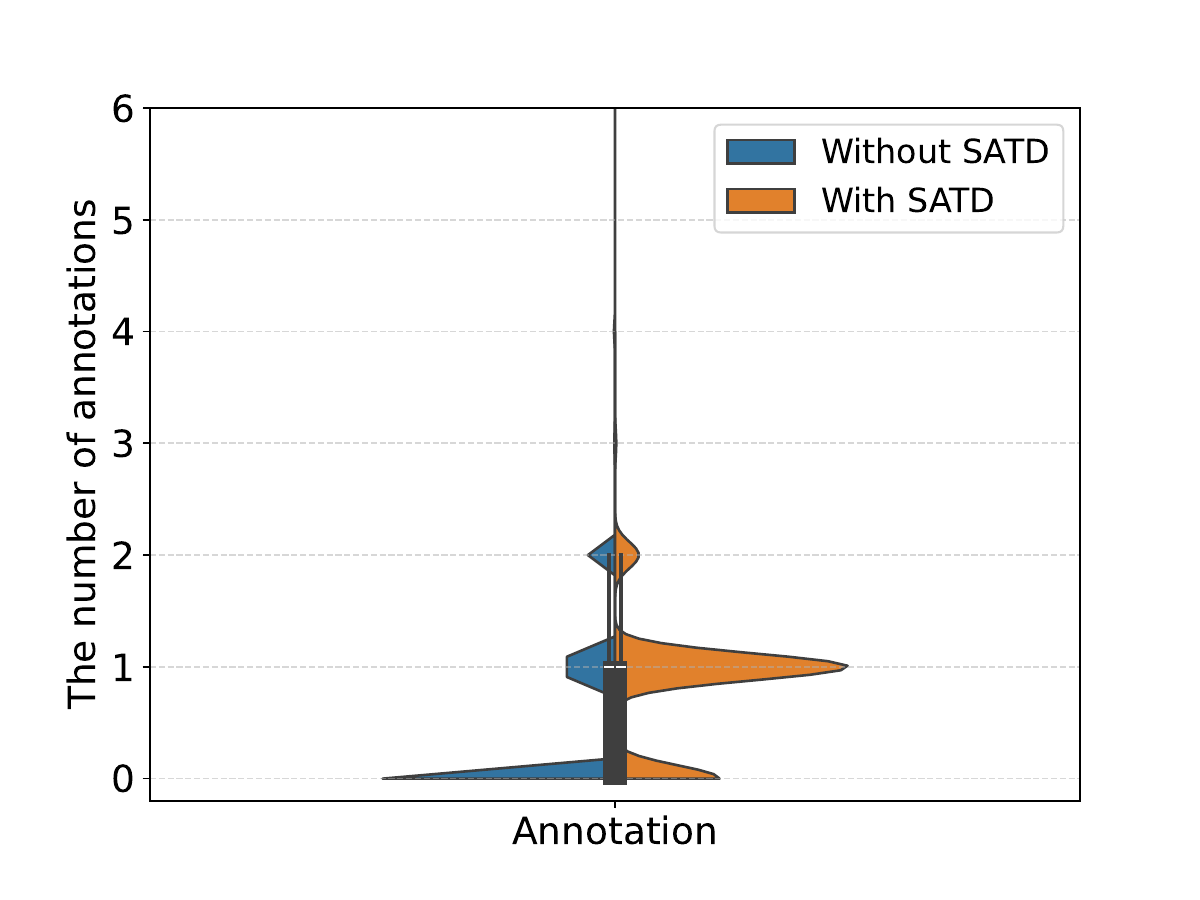}
        \caption{Annotations}
        \label{fig:annotations}
    \end{subfigure}
    \begin{subfigure}{0.45\textwidth}
        \centering
    \end{subfigure}

    \vspace{0.5cm}
    
    \begin{subfigure}{0.45\textwidth}
        \centering
        \includegraphics[width=\linewidth]{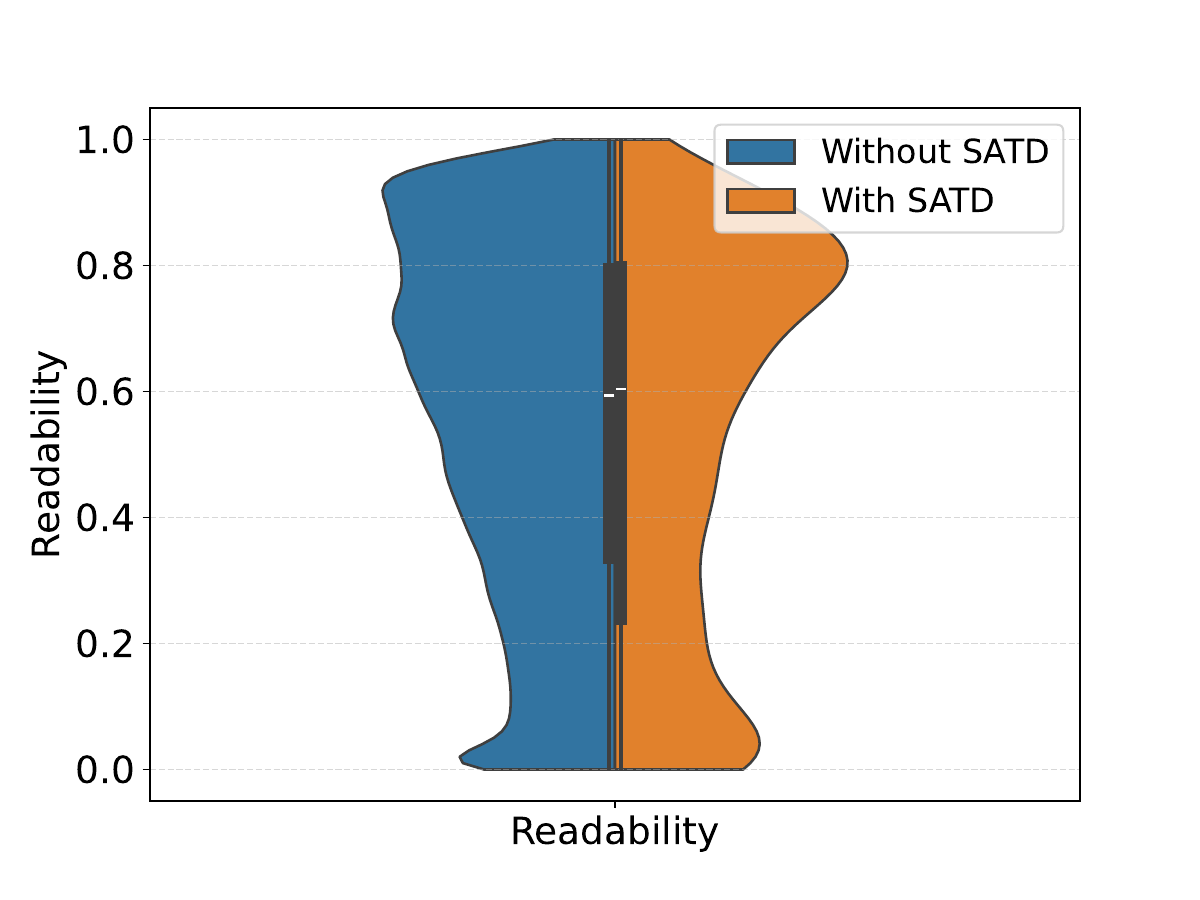}
        \caption{Readability}
        \label{fig:readability}
    \end{subfigure}
    \begin{subfigure}{0.45\textwidth}
        \centering
        \includegraphics[width=\linewidth]{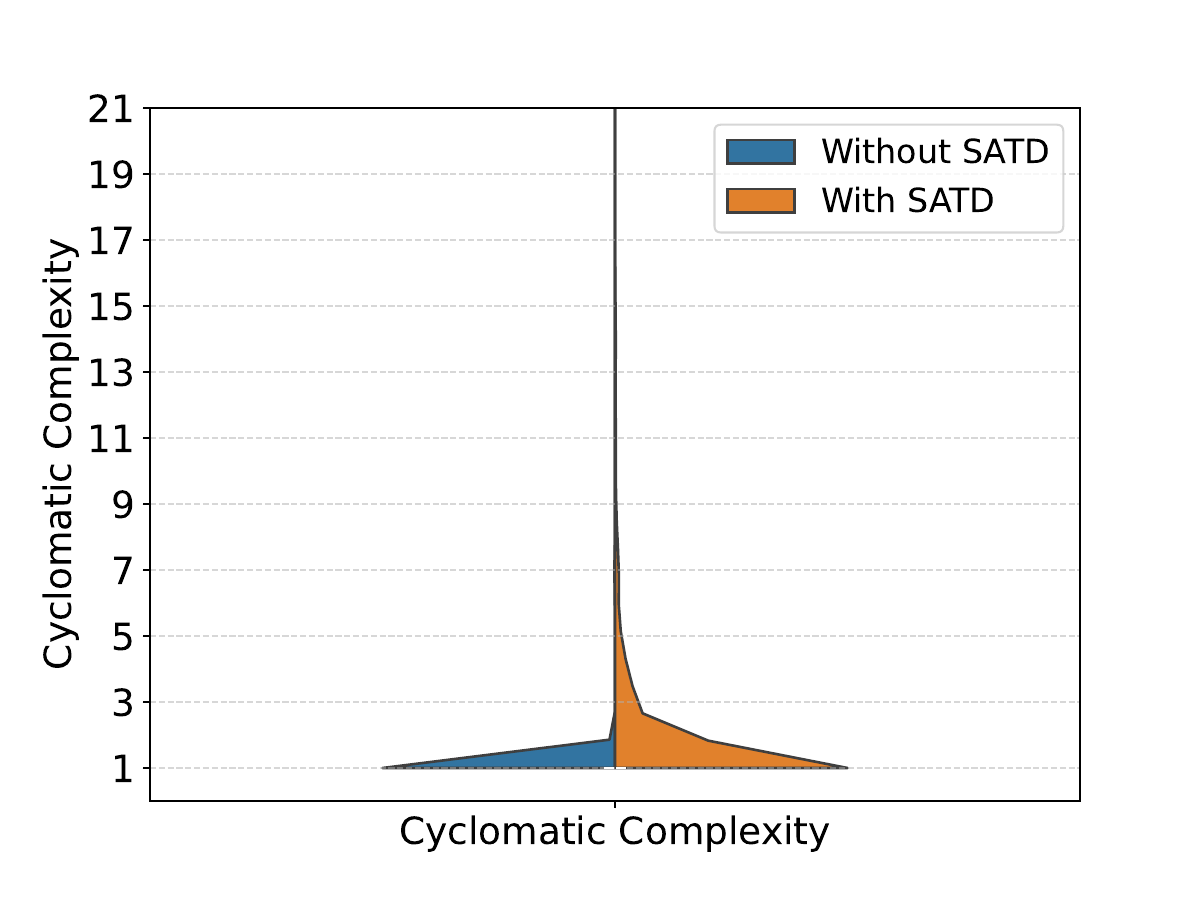}
        \caption{Cyclomatic Complexity}
        \label{fig:cc}
    \end{subfigure}

    \vspace{0.5cm}
\end{figure}

\begin{figure}[htbp]
    \ContinuedFloat
    \centering

    \begin{subfigure}{0.45\textwidth}
        \centering
        \includegraphics[width=\linewidth]{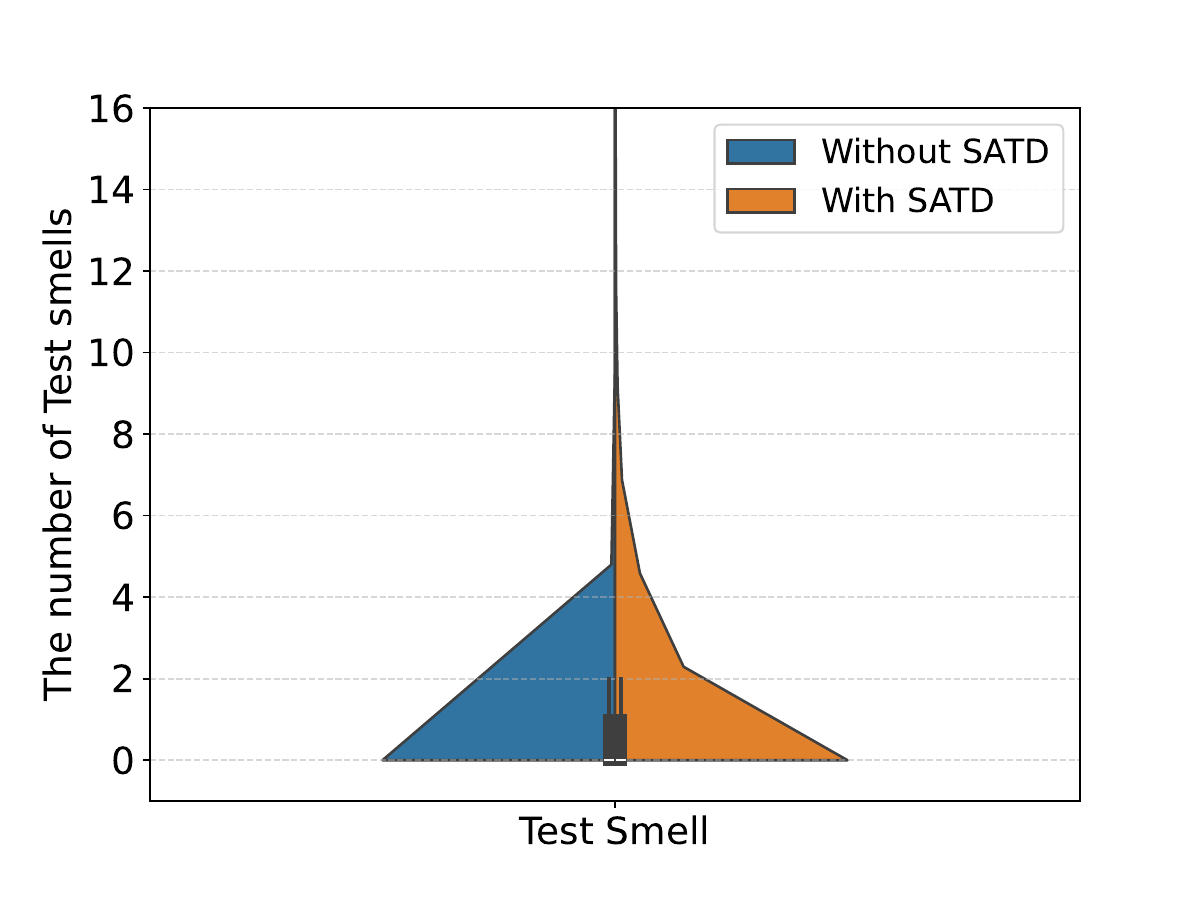}
        \caption{Test smells}
        \label{fig:testsmells}
    \end{subfigure}
    \begin{subfigure}{0.45\textwidth}
        \centering
        \includegraphics[width=\linewidth]{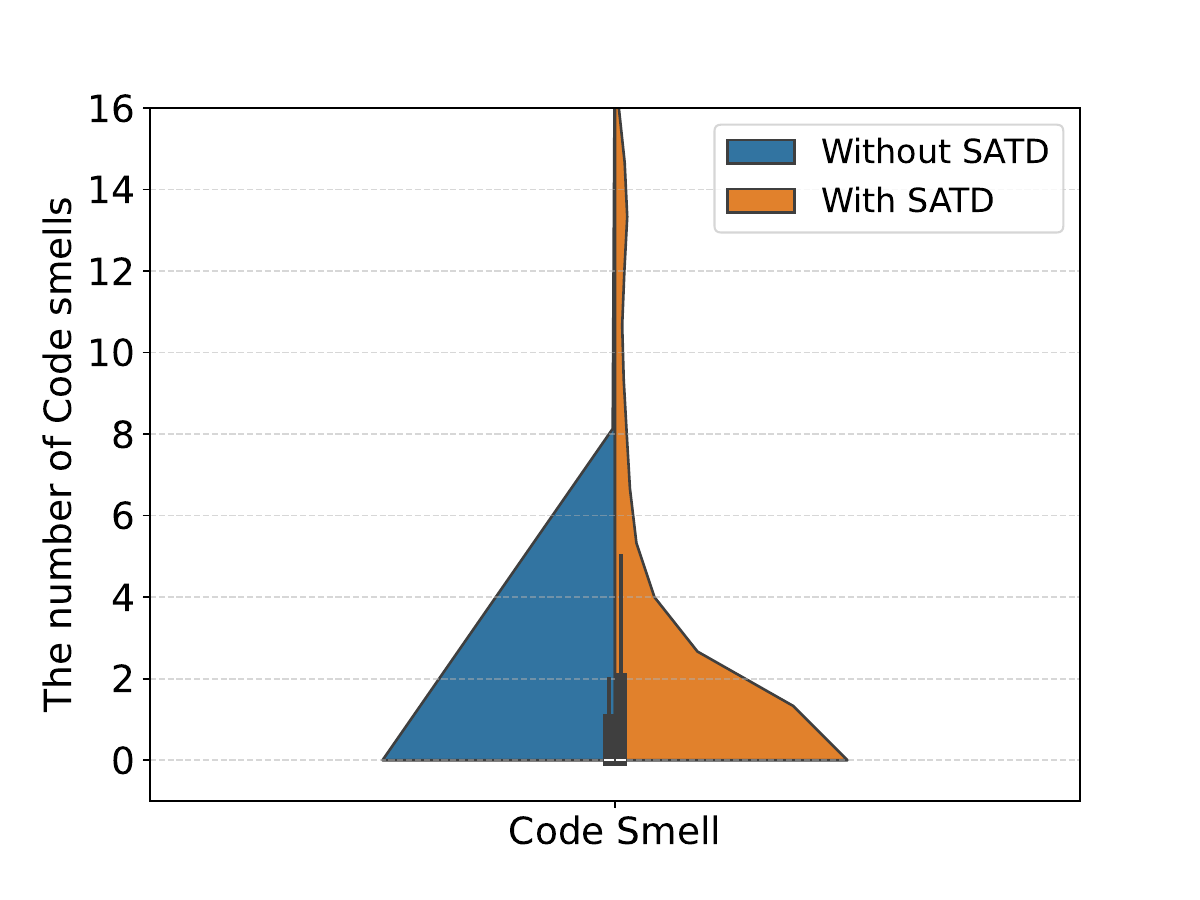}
        \caption{Code smells}
        \label{fig:codesmells}
    \end{subfigure}

    \caption{Distribution in software quality metrics measured in methods with and without \ts}
    \label{fig:distributions}
    \vspace{1cm}
\end{figure}

\begin{figure}[htbp]
    \centering
    \begin{subfigure}{0.45\textwidth}
        \centering
        \includegraphics[width=\linewidth]{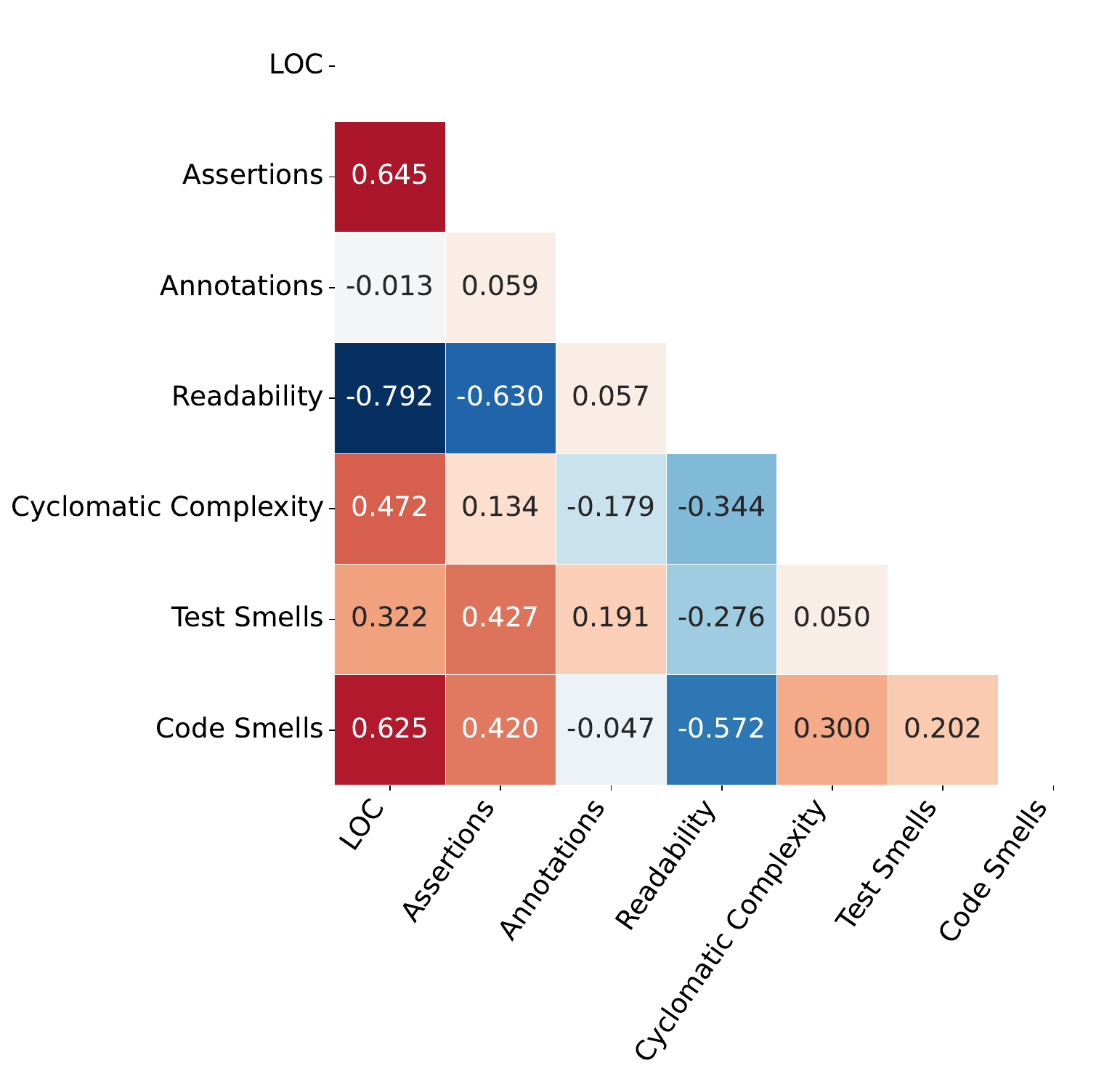}
        \caption{With SATD}
        \label{fig:heatmap-with-satd}
    \end{subfigure}
    \begin{subfigure}{0.45\textwidth}
        \centering
        \includegraphics[width=\linewidth]{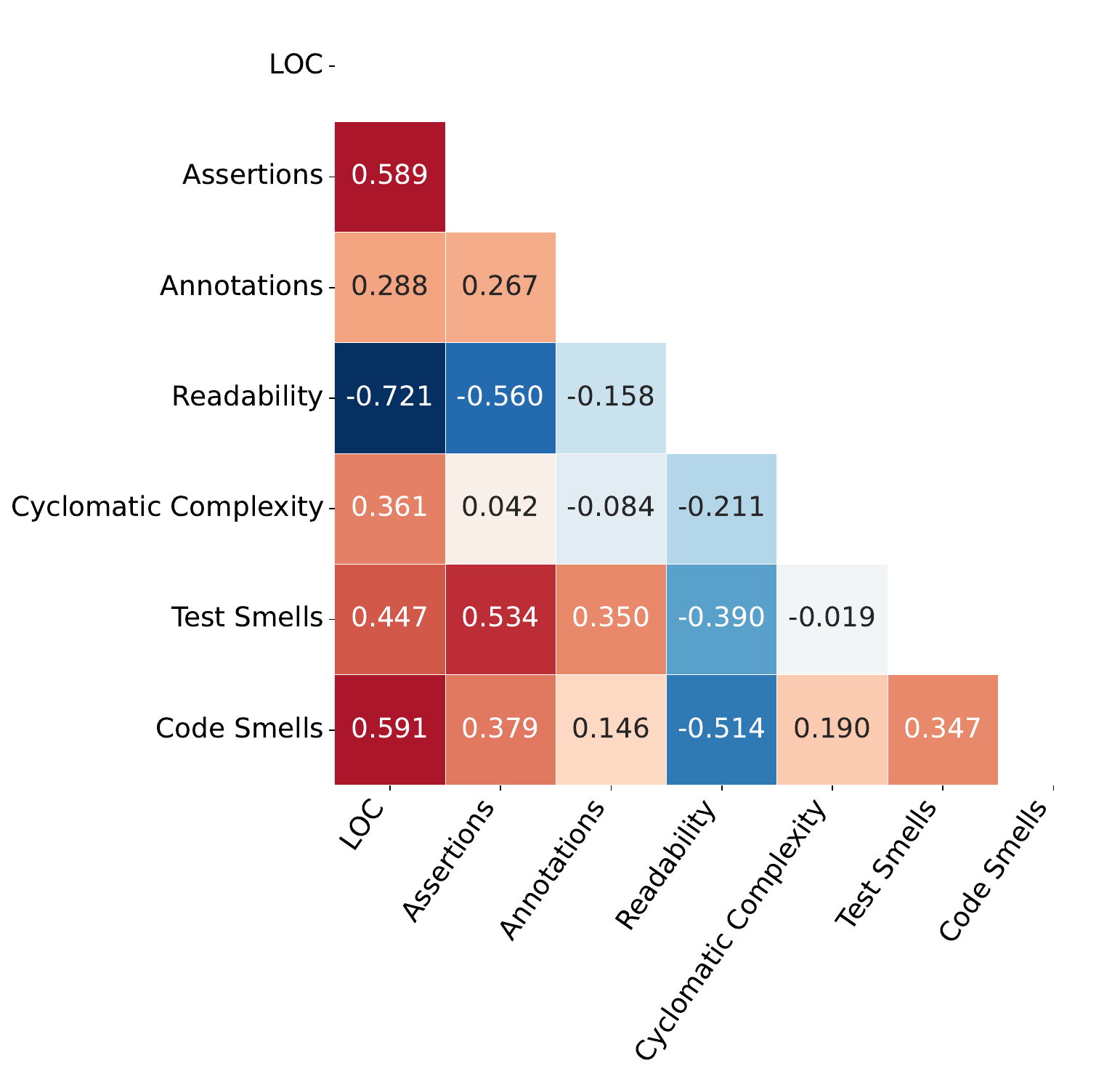}
        \caption{Without SATD}
        \label{fig:heatmap-without-satd}
    \end{subfigure}
    \caption{Spearman Correlation Matrices with and without SATD}
\end{figure}

Furthermore, we conducted a correlation analysis among these quality metrics. We calculated Spearman’s rank correlation coefficient for all pairs of the seven metrics, separately for test methods with and without SATD. 
The Spearman coefficient ($\rho$) ranges from -1 to +1, indicating the direction and strength of a monotonic relationship. A positive value indicates that as one metric increases, the other tends to increase, while a negative value indicates that as one metric increases, the other tends to decrease.
Figure \ref{fig:heatmap-with-satd} and \ref{fig:heatmap-without-satd} visualize these correlation coefficients as two heatmaps.
    
Overall, these two heatmaps present similar correlation trends between the two groups. However, we observed notable differences in the correlations involving the number of annotations. The largest difference was the correlation between Annotations and LOC, which was weakly positive in methods without SATD ($\rho = 0.288$) but near zero in methods with SATD ($\rho = -0.013$), leading to a difference of 0.301. Similarly, the correlation between Annotations and Assertions was also considerably weaker in methods with SATD ($\rho = 0.059$) compared to those without ($\rho = 0.267$), with a difference of 0.208. These findings suggest that the role of annotations may change when developers introduce SATD; in normal test methods, more annotations may be associated with longer code and more assertions, but this relationship appears to break down in the presence of technical debt.


\bigskip
\begin{tcolorbox}
\textbf{RQ2. Test methods with \satd have more lines of code, annotations, code smells, readability, and complexity than those without \satd. However, there is no statistical correlation between Test SATD and test smells, implying that these issues occur independently.}
\end{tcolorbox}

\newpage
\subsection{\rqC}
We inspected \SampleSize randomly selected \satd comments. The process of building the taxonomy, as detailed in Section \ref{sec:rq3-method}, was an iterative effort of labeling and merging. The initial fine-grained labeling of the first batch of 100 instances yielded 146 distinct labels, which were then consolidated into 50 initial categories after a reconciliation meeting. After the full set of \SampleSize instances was classified through the iterative process, a total of 64 categories had been identified. A final round of discussion was then conducted to merge and refine these, resulting in the final taxonomy of \NumberOfDetails categories.
\begin{table*}[t]
\small\centering
\caption{Classification of \ts}
\label{table:RQ3}
\begin{tabular}{ccp{8.4cm}r}
\toprule
Main Cat. & Sub Cat. & Detailed Category & ~~~~~\# \\ \midrule
\addlinespace
\multirow{4}{*}{Production-} & \multirow{2}{*}{Failures} 
& Indicates the failure-triggering environments/settings & \FailuresEnvironment  \\ 
\multirow{4}{*}{originated} &  
& Indicates the failure-triggering inputs                & \FailuresInput        \\ \addlinespace\cline{2-4} 
\addlinespace
\multirow{2.2}{*}{issues} & \multirow{2}{*}{On-hold}  
& Indicates the existence of defects in production code & \OnholdDefect       \\ 
& \multirow{2}{*}{Tasks} 
& Indicates the unimplemented production code        & \OnholdUnimplement  \\ 
&  &  Specifies tasks to execute for debugging          & \OnholdDebug        \\ \addlinespace \midrule
\addlinespace
\multirow{14}{*}{Test-} & \multirow{1}{*}{Test } 
& Indicates incomplete or unimplemented tests             & \IncomleteTest  \\ 
\multirow{14}{*}{originated} & \multirow{0.8}{*}{Completeness} 
& Asks for extra test cases                               & \ExtraTest      \\  \addlinespace\cline{2-4} \addlinespace

\multirow{12.2}{*}{issues} & \multirow{4}{*}{Test Design} 
& Workaround for impossible/difficult implementations               & \Workaround    \\ 
& \multirow{3.8}{*}{and} 
& Indicates non-optimal way of test implementation                  & \NonOptimal    \\ 
& \multirow{3.6}{*}{Implementation} 
& Provides an alternative way to implement/design the test       & \BetterWay     \\ 
&  & Doubts on specific test case design                            & \DoubtsTest     \\
&  & Doubts on test implementation                                  & \DoubtsTestImp \\
&  & Indicates the non-functional issues of tests                   & \NonFunctional \\ \addlinespace\cline{2-4} \addlinespace

& \multirow{6}{*}{Test}  
& Asks for test updates in response to future/latest software versions &  \AskTestUpdate        \\ 
& \multirow{5.8}{*}{Maintenance}  
& Asks for test code restructuring/refactoring                      &  \AskRefactoring       \\ 
&  & Asks for add/update documentation                                       &  \MoreDocument         \\ 
&  & Indicates flaky tests                                             &  \FlakyTest            \\ 
&  & Asks for updates of used production code in tests                 &  \AskProdUpdateInTest  \\ 
&  & Asks for test deletion                                            &  \TestDeletion         \\ 
&  & Indicates invalid or unused tests                                 &  \UnusedTest           \\ \addlinespace \bottomrule
\end{tabular}
\end{table*}
Of the \SampleSize instances, \Matches instances (\MatchRate) are assigned with consistent labels from both reviewers, resulting in a Cohen's kappa coefficient of \Kappa, indicating a substantial agreement according to Landis and Koch~\cite{Landis1977TheMO}. The conflicts were resolved with a third annotator.
Throughout the manual inspection, we found that \FP instances were false positives (incorrectly detected as SATD) and the label of \UnCat instances could not be assigned due to the lack of information. As a result, \ManualSampleSize \satd instances were included in the final results.  
Table \ref{table:RQ3} shows the final classification of \ts and the breakdown of each category. 
In Table \ref{table:RQ3-commonwords}, we identified the top five most frequent words for each category except typical SATD tags such as \texttt{TODO} and \texttt{FIXME}. In particular, we split the words in SATD comments, removed stopwords using the NLTK library,\footnote{\url{https://www.nltk.org}} and identified the most frequent words for each category.
\begin{table}[t]
\centering
\caption{Most frequent words by subcategory}
\label{table:RQ3-commonwords}
\begin{tabular}{ll}
\toprule
Subcategory                   & Common words                              \\ \midrule
Failures                       & \texttt{fails, test, enabled, passes, macos}       \\
On-hold Tasks                  & \texttt{currently, supported, yet, work, fix}      \\
Test Completeness              & \texttt{assert, value, test, check, exception}     \\
Test Design and Implementation & \texttt{test, workaround, need, way, better}       \\
Test Maintenance               & \texttt{remove, version, buffer, response, future} \\ \bottomrule
\end{tabular}
\end{table}

In the following, we introduce each category, separated by the origin of issues (whether the issue originated from production code or test code).

\subsubsection{Production-originated Issues}
This category includes issues intertwined with production code or issues caused by the production code. A total of \ProdIssues \ts instances fall into this category. Production-originated issues are further divided into the following subcategories:

\textbf{Failures: }
This subcategory of SATD indicates the causes of test failures, which are related to either environments/settings or inputs in the production code. 
\Failures cases fell into this category, with \FailuresEnvironment related to environments/settings and \FailuresInput related to specific inputs.
The two most frequently words are \texttt{fails} and \texttt{enabled}, which are related to test execution status.
Bavota\etal\cite{DBLP:conf/msr/BavotaR16} also introduced examples of such failures, but their ``Test debt'' category does not contain any subcategories. Similarly, the study by Kashiwa\etal\cite{DBLP:journals/infsof/KashiwaNKKSSU22} also presents a ``Failure'' category for test-related SATD, but no further categorization was given. We separate the ``failure''-related SATD into two types:

\begin{itemize}
\item \textbf{Indicates the failure-triggering environments/settings:} 
This type of SATD is used to notify developers that a test will fail with a specific environment or configuration. In many cases, developers comment out the assertions that fail or disable the whole test method. 
Snippet~\ref{sni:failures} shows an example, including an access-related failure occurring only on Linux and Windows.\footnote{\url{https://github.com/apache/commons-io/blob/290d72eda9152d1e11b79d48453908ff3f6b9897/src/test/java/org/apache/commons/io/FileUtilsTest.java\#L1730}}
This test method is enabled to run only on Mac OS, using ``\texttt{@EnabledOnOs}'' annotation.\footnote{\url{https://junit.org/junit5/docs/5.2.0/api/org/junit/jupiter/api/condition/EnabledOnOs.html}}

\begin{center}
\captionof{listing}{Example of SATD categorized into ``Indicates the failure-triggering environments/settings''}
\label{sni:failures}
\begin{minted}[highlightlines={2}]{java}
    /**
     * TODO Passes on macOS, fails on Linux and Windows with AccessDeniedException.
     */
    @Test
    @EnabledOnOs(value = OS.MAC)
    public void testForceDeleteUnwritableDirectory() throws Exception {
\end{minted}
\end{center}

\medskip
    \item \textbf{Indicates the failure-triggering inputs:} 
This type of SATD notifies developers that the test will fail when the production code receives a specific input, which is illustrated in Snippet \ref{sni:failureinput}. 
In the example, the test is commented out due to a failure caused by a specific file.\footnote{\url{https://github.com/argouml-tigris-org/argouml/blob/6b6db0242a40f80655cbfdddcca246afe23df20c/src/argouml-core-model-mdr/tests/org/argouml/model/mdr/TestReadCompressedFilesAndHref.java\#L73}}

\begin{center}
\captionof{listing}{Example of SATD categorized into ``Indicates the failure-triggering inputs''}
\label{sni:failureinput}
\begin{minted}[highlightlines={2}]{java}
    public void testReadCompressedAndroMDAProfileIssue5946() {
        // TODO: uncomment the following to get the failure.
//        assertLoadModel(ISSUE5946_BASE_DIR + "zipped-uml14"
//            + "/andromda-profile-datatype/3.3/andromda-profile-datatype-3.3.xml.zip",
//            null, "testReadCompressedAndroMDAProfileFileIssue5946");
    }
\end{minted}

\end{center}

\end{itemize}

\bigskip
\textbf{On-hold Tasks: }
This subcategory of SATD indicates a test is on hold until a specific task in production code is executed, such as new implementation or bug fixing. 
The prevalence of words such as \texttt{currently}, \texttt{yet} shows that developers view these issues as temporary states with an expectation of future resolution.
The ``On-hold SATD'' is originally defined by Maipradit\etal\cite{DBLP:journals/ese/MaipraditTHM20}, which analyzed on-hold SATD in the production code without taking into account the tests. We identified \OnHoldTask cases in this category and the on-hold tasks in test code include the following types:  


\begin{itemize}
\medskip
    \item \textbf{Indicates the existence of defects in production code: } 
This type of SATD is used to inform developers of defects present in the production code. We identified \OnholdDefect instances of this type, making it the most prevalent on-hold task. 
Snippet~\ref{sni:onhold} shows an example of this category, where the test is currently disabled until the defect in the YARN project is fixed.\footnote{\url{https://github.com/apache/hadoop/blob/bd8b77f398f626bb7791783192ee7a5dfaeec760/hadoop-mapreduce-project/hadoop-mapreduce-client/hadoop-mapreduce-client-app/src/test/java/org/apache/hadoop/mapreduce/v2/app/TestStagingCleanup.java\#L224-L227}}

\begin{center}
\captionof{listing}{Example of SATD categorized into ``Indicates the existence of defects in production code''}
\label{sni:onhold}
\begin{minted}[highlightlines={1-4}]{java}
      // FIXME:
      // Disabled this test because currently, when shutdown hook triggered at
      // lastRetry in RM view, cleanup will not do. This should be supported after
      // YARN-2261 completed
    //   @Test (timeout = 30000)
       public void testDeletionofStagingOnKillLastTry() throws IOException {
\end{minted}
\end{center}

\medskip
    \item \textbf{Indicates the unimplemented production code:  }
This type of SATD indicates that a developer is waiting for the implementation of certain production code. 
An example can be seen in Snippet \ref{sni:unimplemented}, in which a method call is commented out because a function for a property is not yet implemented.\footnote{\url{https://github.com/apache/poi/blob/ae2f0945cd2ab37260e46ab46c54b8f68a131aea/poi/src/test/java/org/apache/poi/hpsf/basic/TestWriteWellKnown.java\#L230}}

\begin{center}
\captionof{listing}{Example of SATD categorized into ``Indicates the unimplemented production code''}
\label{sni:unimplemented}
\begin{minted}[highlightlines={2}]{java}
        si.setTemplate(P_TEMPLATE);
        // FIXME (byte array properties not yet implemented): si.setThumbnail(P_THUMBNAIL);
        si.setTitle(P_TITLE);
\end{minted}
\end{center}

\medskip
    \item \textbf{Specifies tasks to execute for debugging: } 
This type of SATD informs developers of the specific task to execute for debugging. 
As illustrated in Snippet \ref{sni:debugging}, 
the SATD comment requests an investigation into why the internal state is not visible during retry processing.\footnote{\url{https://github.com/apache/camel/blob/66736471db8ddc22e50cc31c87d34b072455b488/components/camel-aws/camel-aws-xray/src/test/java/org/apache/camel/component/aws/xray/ErrorTest.java\#L43}}

\begin{center}
\captionof{listing}{Example of SATD categorized into ``Specifies tasks to execute for debugging''}
\label{sni:debugging}
\begin{minted}[highlightlines={1}]{java}
    // FIXME: check why processors invoked in onRedelivery do not generate a subsegment
    public ErrorTest() {
\end{minted}
\end{center}

\end{itemize}

\subsubsection{Test-originated Issues}
This category includes issues originating from the test code. \TestIssues relevant \ts instances are identified, which can be further divided into the following subcategories:

\textbf{Test Completeness: }
This subcategory of SATD pertains to the effectiveness and completeness of test code, such as incomplete test implementations, insufficient test cases, or unclear intent of test cases. 
The category featured words related to test case verification, such as \texttt{check} and \texttt{assert}.
A previous study~\cite{DBLP:journals/infsof/KashiwaNKKSSU22} introduced the ``Necessity'' subcategory, which corresponds to our ``Indicates incomplete or unimplemented tests'' or ``Asks for extra test cases'' categories.
\TestCompleteness cases were identified in this category, with the most frequent one (\IncomleteTest cases) being ``Indicates incomplete or unimplemented tests''. The subcategories in this category are described as follows.

\begin{itemize}
    \item \textbf{Indicates incomplete or unimplemented tests:  }
This type of SATD points out that the test is either incomplete or unimplemented, leading to invalid or ineffective test cases. 
Snippet~\ref{sni:incomplete} shows an example of this subcategory, in which an assertion is needed to check the value produced by the previous line.\footnote{\url{https://github.com/apache/ant/blob/53f19eccf49acf526415997046dca5a5135b0e8f/src/tests/junit/org/apache/tools/ant/IntrospectionHelperTest.java\#L128}}

\begin{center}
\captionof{listing}{Example of SATD categorized into ``Indicates incomplete or unimplemented tests''}
\label{sni:incomplete}
\begin{minted}[highlightlines={4}]{java}
    @Test(expected = BuildException.class)
    public void testElementCreatorTwo() {
        ih.getElementType("two");
        // TODO we should be asserting a value in here
    }
\end{minted}
\end{center}


    \item \textbf{Asks for extra test cases: } 
This type of SATD indicates the need for extra test cases. In contrast to ``Indicates incomplete or unimplemented tests'', this type of SATD requests implementing entire test methods rather than completing existing test methods. They are often created to further ensure the behavior is correct or to improve test coverage. Snippet \ref{sni:extratest} shows an example where an additional test case is required to handle invalid two-byte pairs.\footnote{\url{https://github.com/apache/cassandra/blob/6b134265620d6b39f9771d92edd29abdfd27de6a/test/unit/org/apache/cassandra/db/marshal/TypeValidationTest.java\#L256}}

\begin{center}
\captionof{listing}{Example of SATD categorized into ``Asks for extra test cases''}
\label{sni:extratest}
\begin{minted}[highlightlines={7}]{java}
    @Test
    public void validateUDTNested()
    {
        validate(nestedUDTGen());
    }

    // todo: for completeness, should test invalid two byte pairs.
\end{minted}
\end{center}

\end{itemize}

\bigskip
\textbf{Test Design and Implementation: }
This subcategory of \satd is related to the design and implementation of test code. \TestImplementation cases fall into this category. 
The appearance of words like \texttt{workaround} and \texttt{better} indicates that this category captures SATD related to suboptimal implementations or design improvements needed in tests.
The closest categories from related work are ``Code debt'' and ``Design Debt'' from Bavota\etal's~\cite{DBLP:conf/msr/BavotaR16}. In their subcategories, they also mentioned similar SATD types related to ``workaround'', ``code quality'', and ``design''. 
However, their categories focus on the production code, while we investigate test code. 
In the test code, the relevant SATD can be further categorized as follows: 

\begin{itemize}
    \item \textbf{Indicates workaround for impossible/difficult implementations:} 
This type of SATD is used to inform developers about the workaround for code either impossible or difficult to implement. This is the most frequent SATD type (\Workaround instances) regarding Test Design and Implementation. An example can be seen in Snippet~\ref{sni:workaround1}, in which a developer attempted to retrieve the value of a property, but the current implementation of the production code prevents this due to encapsulation. Consequently, the developer resorted to using reflection to obtain the value. The comment clarifies the rationale behind this implementation choice.\footnote{\url{https://github.com/quarkusio/quarkus/blob/a91a36c533676a5d35ddedbef1717392c9191360/integration-tests/kafka-streams/src/test/java/io/quarkus/it/kafka/streams/KafkaStreamsPropertiesTest.java\#L26}}

\begin{center}
\captionof{listing}{Example of SATD categorized into ``Indicates workaround for impossible/difficult implementations''}
\label{sni:workaround1}
\begin{minted}[highlightlines={3}]{java}
    @Test
    public void testProperties() throws Exception {
        // reflection hack ... no other way to get raw props ...
        Field configField = KafkaStreams.class.getDeclaredField("applicationConfigs");
        configField.setAccessible(true);
        StreamsConfig config = (StreamsConfig) configField.get(streams);
        ...
\end{minted}
\end{center}




    \item \textbf{Indicates non-optimal way of test implementation: } 
This type of SATD is used to highlight non-optimal implementations. Snippet \ref{sni:nonoptimal} provides an example of such a case, where a comment points out the use of a method or approach that is suboptimal. In this instance, the SATD comment indicates that directly using \texttt{DataFormatReifier} is not the best practice and suggests that a better approach should be considered.\footnote{\url{https://github.com/apache/camel/blob/66736471db8ddc22e50cc31c87d34b072455b488/components/camel-fhir/camel-fhir-component/src/test/java/org/apache/camel/component/fhir/dataformat/spring/FhirDataformatConfigSpringTest.java\#L90}}

\begin{center}
\captionof{listing}{Example of SATD categorized into ``Indicates non-optimal way of test implementation''}
\label{sni:nonoptimal}
\begin{minted}[highlightlines={3}]{java}
    private FhirDataFormat getDataformat(String name) {
        CamelContext camelContext = context();
        // TODO: Do not use reifier directly
        return (FhirDataFormat) DataFormatReifier.getDataFormat(camelContext, name);
    }
\end{minted}
\end{center}
\newpage
    \item \textbf{Provides an alternative way to implement/design the test:}  
This type of SATD is used to propose a different way to implement the test. Snippet \ref{sni:betterway} provides an example of this type of SATD, where the comment suggests using ``interrupts'' instead of \texttt{notifyAll()} to target waiting threads more effectively.\footnote{\url{https://github.com/apache/cassandra/blob/6b134265620d6b39f9771d92edd29abdfd27de6a/test/simulator/main/org/apache/cassandra/simulator/systems/InterceptingMonitors.java\#L363}}

\begin{center}
\captionof{listing}{Example of SATD categorized into ``Provides an alternative way to implement/design the test''}
\label{sni:betterway}
\begin{minted}[highlightlines={4}]{java}
    if (waitingOnRelinquish)
    {
        waitingOnRelinquish = false;
        monitor.notifyAll(); // TODO: could use interrupts to target waiting anyway, avoiding notifyAll()
    }
\end{minted}
\end{center}

    \item \textbf{Doubts on specific test case design: } 
This type of SATD raises questions about purposes, validity or effectiveness of test cases.  
Snippet~\ref{sni:doubttestcase} provides an example where a developer questioned the purpose of the test case and subsequently commented out the entire test method.\footnote{\url{https://github.com/apache/bookkeeper/blob/cce4b6461691466c663f2cb4d00dd4d73dd9071e/bookkeeper-server/src/test/java/org/apache/bookkeeper/bookie/datainteg/DataIntegrityCheckTest.java\#L410}}

\begin{center}
\captionof{listing}{Example of SATD categorized into ``Doubts on specific test case design''}
\label{sni:doubttestcase}
\begin{minted}[highlightlines={1}]{java}
    // TODO: what is this test?
//    @Test
//    public void testRecoverLimboFlushFailure() throws Exception {
//        MockLedgerManager lm = new MockLedgerManager();
          ...
\end{minted}
\end{center}


    \item \textbf{Doubts on test implementation:}  
This type of SATD appears when developers are unclear about whether the implementations are optimal or not. Snippet \ref{sni:doubt-implementation} provides an example, in which the comment raises questions about the necessity of the \texttt{Thread.sleep()} call, suggesting that a further review is needed.\footnote{\url{https://github.com/apache/camel/blob/66736471db8ddc22e50cc31c87d34b072455b488/components/camel-github/src/test/java/org/apache/camel/component/github/consumer/PullRequestCommentConsumerTest.java\#L55}}

\begin{center}
\captionof{listing}{Example of SATD categorized into ``Doubts on test implementation''}
\label{sni:doubt-implementation}
\begin{minted}[highlightlines={3}]{java}
    mockResultEndpoint.expectedBodiesReceivedInAnyOrder(commitComment1, commitComment2);

    Thread.sleep(1 * 1000);         // TODO do I need this?

    mockResultEndpoint.assertIsSatisfied();
\end{minted}

\end{center}

\smallskip
    \item \textbf{Indicates the non-functional issues of tests:} 
This type of SATD is used to inform developers about non-functional issues in the test code, such as performance, scalability, or resource usage. Snippet \ref{sni:nonfunctional} provides an example, where the comment indicates that the test takes an excessive amount of time to execute, potentially impacting overall test efficiency.\footnote{\url{https://github.com/cloudfoundry/uaa/blob/5b74878c0297043860bb88434ee1123934acfe19/server/src/test/java/org/cloudfoundry/identity/uaa/util/CachingPasswordEncoderTest.java\#L125}}

\begin{center}
\captionof{listing}{Example of SATD categorized into ``Indicates the non-functional issues of tests''}
\label{sni:nonfunctional}
\begin{minted}[highlightlines={2}]{java}
    @Test
    // TODO: This test takes a long time to run :(
    void ensureNoMemoryLeak() {
\end{minted}
\end{center}


\end{itemize}

\bigskip
\textbf{Test Maintenance: }
This subcategory of \satd is related to test code maintenance. These \satd instances indicate the need for updates in future versions, refactoring, fixing flaky tests, and adding documentation. 
Frequent terms such as \texttt{remove} and \texttt{version} indicate that this category captures SATD related to test code management, including cleanup and version updates.
This subcategory does not exist in Bavota\etal's work~\cite{DBLP:conf/msr/BavotaR16}, but ``Asks for test updates in response to future/latest software versions'' partially matches the ``Future work'' subcategory in Kashiwa\etal's study~\cite{DBLP:journals/infsof/KashiwaNKKSSU22}. We identified \TestMaintenance cases in this category, with the most frequent subcategory being ``Asks for test updates in response to future/latest software versions,'' represented by \AskTestUpdate instances. The details of each subcategory are described below.


\begin{itemize}

\medskip
    \item \textbf{Asks for test updates in response to future/latest software versions:  } 
This type of SATD is used to inform developers that the test needs to be updated in a future version of production code. Snippet~\ref{sni:askupdate1} provides an example of this subcategory. The comment suggests the removal of the reference to SecurityManager in this test case when the project updates the version of the programming language Groovy, as the class is discontinued in the new version.\footnote{\url{https://github.com/apache/groovy/blob/f6221ee780bfb2f84fb197da2c13387c4e93a019/src/test/org/codehaus/groovy/reflection/SecurityTest.java\#L301}}

\begin{center}
\captionof{listing}{Example of SATD categorized into ``Asks for test updates in response to future/latest software versions''}
\label{sni:askupdate1}
\begin{minted}[highlightlines={1}]{java}
    @SuppressWarnings("removal") // TODO in a future Groovy version remove reference to SecurityManager, for now not run for JDK18+
    public void testInvokesPrivateMethodsInGroovyObjectsWithoutChecks() throws Exception {
        if (isAtLeastJdk("18.0")) return;
        ...
\end{minted}
\end{center}


\medskip
    \item \textbf{Asks for test code restructuring/refactoring}: 
This type of SATD is used to notify developers that the test code needs to be restructured or refactored.
Snippet \ref{sni:refactoring} shows an example, where the SATD comment suggests making the contract more explicit or extracting common code to improve readability and maintainability.\footnote{\url{https://github.com/apache/flink/blob/eaffd227d853e0cdef03f1af5016e00f950930a9/flink-state-backends/flink-statebackend-changelog/src/test/java/org/apache/flink/state/changelog/ChangelogStateDiscardTest.java\#L376}}
When looking into 9 cases, most of the SATDs request structural refactoring, including five instances of \textsc{Move Method/Class}, two of \textsc{Extract Method}, one of \textsc{Rename Method}, and one involving data structure changes (\eg Introduce Parameter Object). We also found that only one case co-occurred with a test smell (specifically, ``Assertion Roulette''), reinforcing our finding that Test SATD and test smells are largely independent issues.
\newpage
\begin{center}
\captionof{listing}{Example of SATD categorized into ``Asks for test code restructuring / refactoring''}
\label{sni:refactoring}
\begin{minted}[highlightlines={1}]{java}
            // todo: make the contract more explicit or extract common code
            Map<UploadTask, Map<StateChangeSet, Tuple2<Long, Long>>> taskOffsets =
                    tasks.stream().collect(toMap(identity(), this::mapOffsets));
\end{minted}
\end{center}


    \item \textbf{Asks for add/update documentation:}  
This type of SATD is used to request additional documentation or updates to the documentation to understand the goals and implementation of the test code. Snippet \ref{sni:documentation} provides an example, where a comment highlights the need for documentation to clarify specific behavior.\footnote{\url{https://github.com/spring-projects/spring-framework/blob/f85d5bd84a7e7bc810bb5a8179fc2fc130affc89/spring-aop/src/test/java/org/springframework/aop/aspectj/annotation/AbstractAspectJAdvisorFactoryTests.java\#L435}}

\begin{center}
\captionof{listing}{Example of SATD categorized into ``Asks for add/update documentation''}
\label{sni:documentation}
\begin{minted}[highlightlines={1}]{java}
    // TODO document this behaviour.
    // Is it different AspectJ behaviour, at least for checked exceptions?
    @Test
    void aspectMethodThrowsExceptionIllegalOnSignature() {
        TestBean target = new TestBean();
        RemoteException expectedException = new RemoteException();
        List<Advisor> advisors = getAdvisorFactory().getAdvisors(
            aspectInstanceFactory(new ExceptionThrowingAspect(expectedException), "someBean"));
        assertThat(advisors).as("One advice method was found").hasSize(1);
        ITestBean itb = createProxy(target, ITestBean.class, advisors);
        assertThatExceptionOfType(UndeclaredThrowableException.class)
            .isThrownBy(itb::getAge)
            .withCause(expectedException);
    }
\end{minted}
\end{center}

    \item \textbf{Indicates flaky tests:}  
This type of SATD informs developers that the test is unstable—sometimes passing and sometimes failing (\ie flaky tests)~\cite{DBLP:conf/sigsoft/EckPCB19}. Snippet \ref{sni:flaky} provides an example of such SATD. In this instance, the comment warns that the implementation causes flaky tests.\footnote{\url{https://github.com/apache/commons-io/blob/290d72eda9152d1e11b79d48453908ff3f6b9897/src/test/java/org/apache/commons/io/FileUtilsWaitForTest.java\#L42}}

\begin{center}
\captionof{listing}{Example of SATD categorized into ``Indicates flaky tests''}
\label{sni:flaky}
\begin{minted}[highlightlines={2}]{java}
    /**
     * TODO Fails randomly.
     */
    @Test
    public void testWaitForInterrupted() throws InterruptedException {
\end{minted}
\end{center}

    \item \textbf{Asks for updates of used production code in tests:}  
This type of SATD is used to indicate outdated production code used in the test. Snippet \ref{sni:askupdateofusedproduction} provides an example, where a comment requests updating the test to use a new conflict resolver.\footnote{\url{https://github.com/apache/maven/blob/8b094c9513efc1b9ce2d952b3b9c8eaedaf8cbf0/maven-compat/src/test/java/org/apache/maven/repository/legacy/resolver/DefaultArtifactCollectorTest.java\#L155}}

\begin{center}
\captionof{listing}{Example of SATD categorized into ``Asks for updates of used production code in tests''}
\label{sni:askupdateofusedproduction}
\begin{minted}[highlightlines={3}]{java}
    public void disabledtestResolveCorrectDependenciesWhenDifferentDependenciesOnNewest()
            throws ArtifactResolutionException, InvalidVersionSpecificationException {
        // TODO use newest conflict resolver
        ArtifactSpec a = createArtifactSpec("a", "1.0");
        ArtifactSpec b = a.addDependency("b", "1.0");
        ArtifactSpec c2 = b.addDependency("c", "2.0");
        ArtifactSpec d = c2.addDependency("d", "1.0");
\end{minted}
\end{center}

    \item \textbf{Asks for test deletion: }
This type of SATD is used to report unnecessary tests. Snippet \ref{sni:deletion} provides an example, where a comment requests the deletion of a test that is no longer needed.\footnote{\url{https://github.com/apache/dubbo/blob/3609ddb2259ad223f6c0a827e36f6f8ccd38c6b2/dubbo-config/dubbo-config-api/src/test/java/org/apache/dubbo/config/MethodConfigTest.java\#L111}}

\begin{center}
\captionof{listing}{Example of SATD categorized into ``Asks for test deletion''}
\label{sni:deletion}
\begin{minted}[highlightlines={1}]{java}
    // TODO remove this test
    @Test
    void testStaticConstructor() throws NoSuchFieldException {
\end{minted}
\end{center}

    \item \textbf{Indicates invalid or unused tests: } 
This type of SATD is used to inform other developers that the test is currently invalid or unused. Snippet \ref{sni:unused} provides an example of such SATD, where the comment asks other developers to verify if the test is unused.\footnote{\url{https://github.com/argouml-tigris-org/argouml/blob/6b6db0242a40f80655cbfdddcca246afe23df20c/src/argouml-app/tests/org/argouml/model/TestAgainstUmlModel.java\#L91}}

\begin{center}
\captionof{listing}{Example of SATD categorized into ``Indicates invalid or unused tests''}
\label{sni:unused}
\begin{minted}[highlightlines={6}]{java}
    /**
     * @throws SAXException when things go wrong with SAX
     * @throws IOException when there's an IO error
     * @throws ParserConfigurationException when the parser finds wrong syntax
     * 
     * TODO: Unused?
     */
    public void testDataModel()
	throws SAXException,
	       IOException,
	       ParserConfigurationException {
\end{minted}
\end{center}
    
\end{itemize}


\bigskip
\begin{tcolorbox}
\textbf{RQ3.
Test SATD serves various purposes and can be classified into five major categories related to \NumberOfDetails types of issues. The most frequent purpose of \ts is ``Indicating incomplete or unimplemented tests''. Additionally, most of the issues identified in this study do not fit the existing categories proposed by previous studies, highlighting the differences between SATD in the production and test code.
}
\end{tcolorbox}

\newpage
\subsection{\rqD}
Table \ref{table:RQ4_svm} presents the results of classifying \ts into five subcategories using machine-learning and deep-learning classifiers.
First, in terms of precision, the GPT-3.5-turbo model achieved the highest overall precision (\ie 0.77), indicating its effectiveness in preventing false positives. Conversely, the XGBoost classifier exhibited the lowest overall precision, with a value of 0.59.

When looking into the recall, deep-learning models overall outperformed traditional machine-learning models. The highest recall of 0.67 was jointly achieved by the fine-tuned CodeBERT model and the large language model GPT-4.1. The BERT-based classifier also performed well with a recall of 0.64.
In contrast, the Naive Bayes classifier recorded the lowest recall at 0.39, resulting in a higher number of undetected instances.

In terms of F1-score, the CodeBERT-based classifier achieved the highest overall score (\ie 0.70), demonstrating robustness in accurately identifying relevant SATD instances while minimizing misclassifications. Notably, the superiority of CodeBERT over BERT suggests that characteristic words specific to test code are more prevalent than in natural language, a finding also observed in studies focusing on production code~\cite{DBLP:journals/ese/SheikhaeiTWX24}.

\begin{figure}[t]
    \centering
    \includegraphics[width=0.8\linewidth]{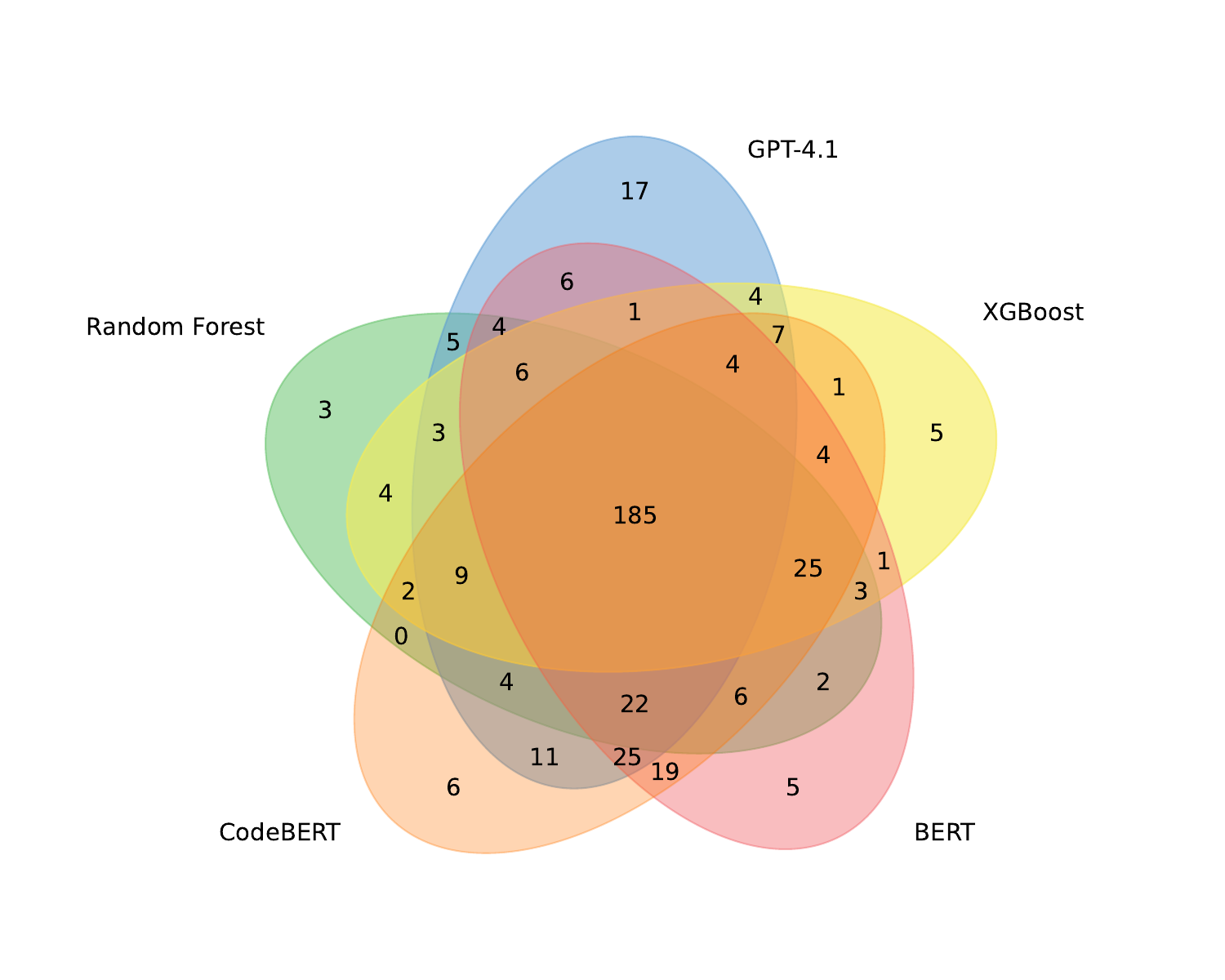}
    \caption{Venn diagram of correct predictions for the top five classification models}
    \label{fig:venn_diagram}
\end{figure}

Figure \ref{fig:venn_diagram} presents a Venn diagram illustrating the distribution of correctly classified instances among the top five classifiers by F1-score.\footnote{Our Venn diagram does not include results from all eight classifiers as such plots are unreadable. Instead, plots with five classifiers are easier to comprehend.} 
185 out of 430 instances were correctly predicted by all the classifiers. Furthermore, this qualitative analysis highlights a crucial finding that the F1-score alone does not capture: the GPT-4.1 model made the most uniquely correct predictions (17 instances), surpassing all other models, including the top-performing CodeBERT. This strongly suggests that while CodeBERT is a robust, well-balanced classifier, modern LLMs like GPT-4.1 possess a distinct capability to understand different semantic nuances in SATD comments that other models may miss.

Next, looking into the prediction performance for subcategories, the best-performing model CodeBERT exhibits outstanding performance for predicting SATDs falling into ``Test Completeness'' category (\ie F1 score of 0.83). On the other hand, the lowest F1-score was observed in the ``Failures'' category with a value of 0.54. 
In particular, instances of the ``Failures'' category were frequently misclassified as ``On-hold Task''. For example, we observe several cases that have a trigger condition which is often used in ``On-hold Task'' category such as ``\texttt{TODO: WriteResult isn't returned when inserting}''.\footnote{\url{https://github.com/apache/camel/blob/66736471db8ddc22e50cc31c87d34b072455b488/components/camel-mongodb/src/test/java/org/apache/camel/component/mongodb/integration/MongoDbHeaderHandlingIT.java\#L71}} This low performance is likely to be caused by the lack of instances (\ie the dataset contains only 15 instances from the ``Failures'' category). Our future work will extend the dataset to increase the number of SATD that fall into these smaller categories. 


\bigskip
\begin{tcolorbox}
\textbf{RQ4.
The CodeBERT-based model outperforms other machine learning models in terms of Recall and F1-Score (0.67 and 0.70, respectively). It achieves the highest performance in the category of ``Test Completeness'' (0.83), while showing the lowest performance in the ``Failures'' category (0.54). While the prediction results are reasonable, there is still large room for improvement.
}
\end{tcolorbox}

\begin{table}[t]
\footnotesize
\centering
\caption{Performance comparison using different machine-learning or deep-learning algorithms}
\label{table:RQ4_svm}
\subcaption{SVM (Accuracy: 0.63)}
\begin{tabular}{rrrrrr|r}
\toprule
          & Failures & On-hold Task & \begin{tabular}[c]{@{}l@{}}Test  Completeness\end{tabular} & \begin{tabular}[c]{@{}l@{}}Test Design\\  and Implementation\end{tabular}  & Test Maintenance & Average\\ \midrule
Precision & 0.83     & 0.71   & 0.73 & 0.54    & 1.00 & 0.76\\
Recall    & 0.33     & 0.36   & 0.61 & 0.87    & 0.31 & 0.50\\
F1-Score  & 0.48     & 0.48   & 0.67 & 0.67    & 0.48 & 0.55\\ \bottomrule
\end{tabular}

\vspace{0.5cm}
\label{table:RQ4_naivebayes}
\subcaption{Naive Bayes (Accuracy: 0.59)}
\begin{tabular}{rrrrrr|r}
\toprule
          & Failures & On-hold Task & \begin{tabular}[c]{@{}l@{}}Test  Completeness\end{tabular} & \begin{tabular}[c]{@{}l@{}}Test Design\\  and Implementation\end{tabular}  & Test Maintenance & Average\\ \midrule
Precision & 0.00  & 0.87  & 0.57 & 0.56    & 1.00 & 0.60 \\
Recall    & 0.00  & 0.19  & 0.73 & 0.78    & 0.25 & 0.39 \\
F1-Score  & 0.00  & 0.32  & 0.64 & 0.65    & 0.40 & 0.40 \\ \bottomrule
\end{tabular}

\vspace{0.5cm}
\label{table:RQ4_xgboost}
\subcaption{XGBoost (Accuracy: 0.61)}
\begin{tabular}{rrrrrr|r}
\toprule
          & Failures & On-hold Task & \begin{tabular}[c]{@{}l@{}}Test  Completeness\end{tabular} & \begin{tabular}[c]{@{}l@{}}Test Design\\  and Implementation\end{tabular}  & Test Maintenance & Average\\ \midrule
Precision & 0.54  & 0.61  & 0.64 & 0.61    & 0.54 & 0.59 \\
Recall    & 0.47  & 0.52  & 0.63 & 0.71    & 0.42 & 0.55 \\
F1-Score  & 0.50  & 0.56  & 0.63 & 0.66    & 0.47 & 0.57 \\ \bottomrule
\end{tabular}

\vspace{0.5cm}
\label{table:RQ4_randomforest}
\subcaption{Random Forest (Accuracy: 0.66)}
\begin{tabular}{rrrrrr|r}
\toprule
          & Failures & On-hold Task & \begin{tabular}[c]{@{}l@{}}Test  Completeness\end{tabular} & \begin{tabular}[c]{@{}l@{}}Test Design\\  and Implementation\end{tabular}  & Test Maintenance & Average\\ \midrule
Precision & 0.67  & 0.76  & 0.73 & 0.58    & 0.91 & 0.73 \\
Recall    & 0.27  & 0.37  & 0.69 & 0.85    & 0.42 & 0.52 \\
F1-Score  & 0.38  & 0.50  & 0.71 & 0.69    & 0.57 & 0.57 \\ \bottomrule
\end{tabular}

\vspace{0.5cm}
\label{table:RQ4_bert}
\subcaption{BERT (Accuracy: 0.74)}
\begin{tabular}{rrrrrr|r}
\toprule
          & Failures & On-hold Task & \begin{tabular}[c]{@{}l@{}}Test  Completeness\end{tabular} & \begin{tabular}[c]{@{}l@{}}Test Design\\  and Implementation\end{tabular}  & Test Maintenance & Average\\ \midrule
Precision & 0.55  & 0.59  & 0.83 & 0.74    & 0.73 & 0.69 \\
Recall    & 0.40  & 0.54  & 0.79 & 0.85    & 0.62 & 0.64 \\
F1-Score  & 0.46  & 0.56  & 0.81 & 0.79    & 0.67 & 0.66 \\ \bottomrule
\end{tabular}

\vspace{0.5cm}
\label{table:RQ4_codebert}
\subcaption{CodeBERT (Accuracy: \bf{0.77})}
\begin{tabular}{rrrrrr|r}
\toprule
          & Failures & On-hold Task & \begin{tabular}[c]{@{}l@{}}Test  Completeness\end{tabular} & \begin{tabular}[c]{@{}l@{}}Test Design\\  and Implementation\end{tabular}  & Test Maintenance & Average\\ \midrule
Precision & 0.64  & 0.65  & 0.85 & 0.77    & 0.72 & 0.73 \\
Recall    & 0.47  & 0.63  & 0.82 & 0.86    & 0.60 & \bf{0.67} \\
F1-Score  & 0.54  & 0.64  & 0.83 & 0.81    & 0.66 & \bf{0.70} \\ \bottomrule
\end{tabular}

\end{table}

\begin{table}[t]
\ContinuedFloat
\footnotesize
\centering
\vspace{0.5cm}
\label{table:RQ4_gpt-3.5}
\subcaption{GPT-3.5-turbo (Accuracy: 0.59)}
\begin{tabular}{rrrrrr|r}
\toprule
          & Failures & On-hold Task & \begin{tabular}[c]{@{}l@{}}Test  Completeness\end{tabular} & \begin{tabular}[c]{@{}l@{}}Test Design\\  and Implementation\end{tabular}  & Test Maintenance & Average\\ \midrule
Precision & 1.00  & 1.00  & 0.93 & 0.55    & 0.41 & \bf{0.77} \\
Recall    & 0.40  & 0.07  & 0.45 & 0.91    & 0.69 & 0.50 \\
F1-Score  & 0.57  & 0.14  & 0.61 & 0.69    & 0.51 & 0.50 \\ \bottomrule
\end{tabular}
\vspace{0.5cm}
\label{table:RQ4_gpt-4.1}
\subcaption{GPT-4.1 (Accuracy: 0.73)}
\begin{tabular}{rrrrrr|r}
\toprule
          & Failures & On-hold Task & \begin{tabular}[c]{@{}l@{}}Test  Completeness\end{tabular} & \begin{tabular}[c]{@{}l@{}}Test Design\\  and Implementation\end{tabular}  & Test Maintenance & Average\\ \midrule
Precision & 0.45  & 0.61  & 0.95 & 0.71    & 0.58 & 0.66 \\
Recall    & 0.60  & 0.30  & 0.72 & 0.89    & 0.83 & \bf{0.67} \\
F1-Score  & 0.51  & 0.40  & 0.82 & 0.79    & 0.68 & 0.64 \\ \bottomrule
\end{tabular}
\end{table}

\clearpage
\section{Discussions}\label{sec:discussions}
\subsection{Test Smells with Types of Test SATD}
RQ2 examined the correlation between SATD in test code and various test quality indicators, including general code quality metrics (like code smells, lines of code, complexity, and readability) and test-specific metrics (such as test smells, assertions, and annotations) at the method level. RQ3 investigated the purposes and types of Test SATD, leading to a detailed taxonomy of 20 issue types categorized into five main groups.

While RQ2 found no direct statistical correlation between the general presence of Test SATD and test smells, it was not clear whether specific types of Test SATD identified in RQ3 might impact test quality, particularly regarding test smells.
To examine this further, an additional analysis was conducted to study the impact of each type of SATD on quality aspects. This analysis leveraged the CodeBERT-based classifier, which was developed in RQ4 and demonstrated the highest overall performance (F1-score of 0.70) among the evaluated models for automatically classifying Test SATD types. This classifier was applied to all 2,092 filtered SATD instances to enhance the reliability and comprehensiveness of the analysis.

\begin{figure}[t]
    \centering
    \includegraphics[width=0.58\linewidth]{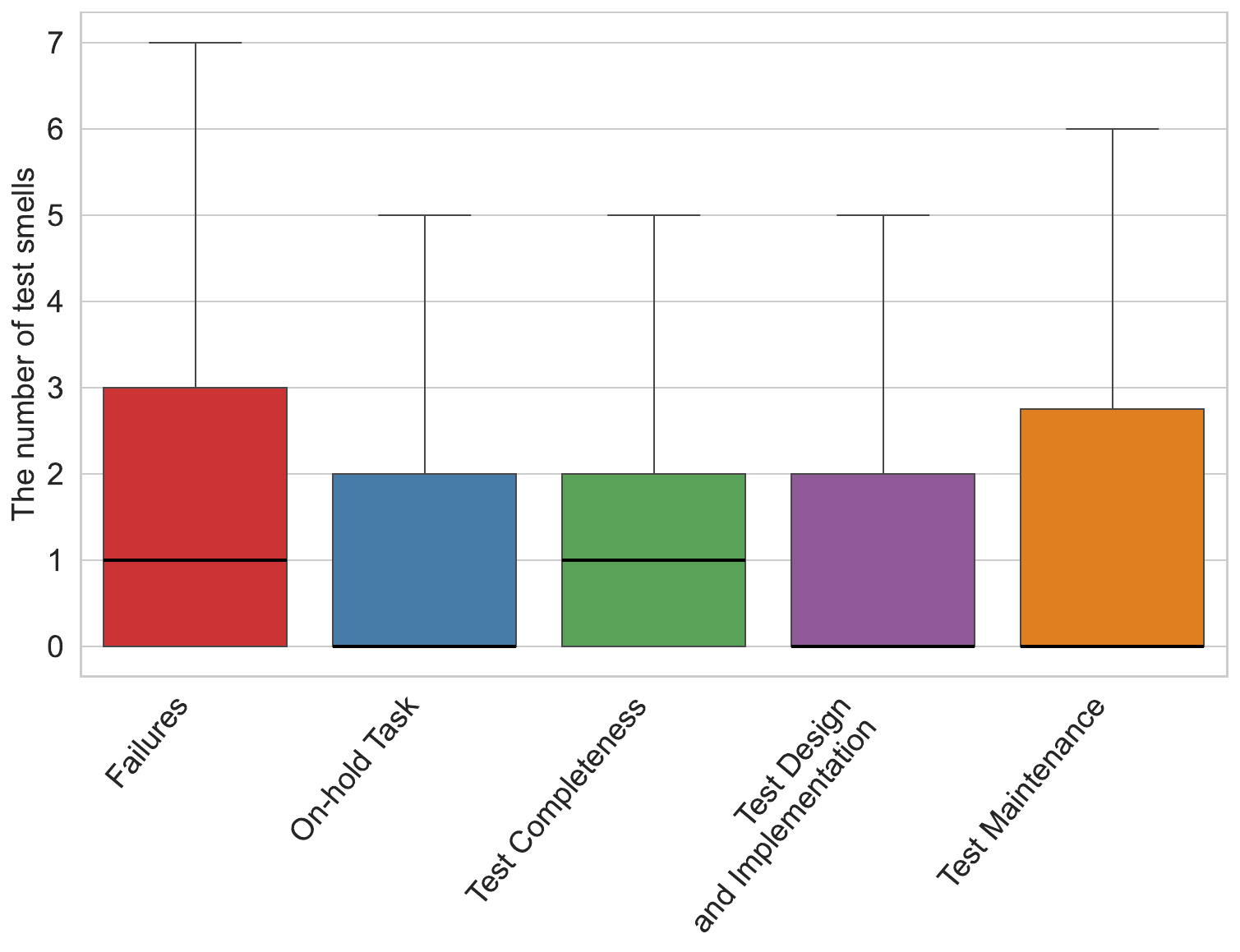}
    \caption{The number of test smells by category}
    \label{fig:testsmells-by-category}
\end{figure}

Figure \ref{fig:testsmells-by-category} shows the distribution of the number of test smells across the SATD subcategories. Focusing on the median values, we observed that the ``Test Completeness'' and ``Failures'' categories had a median of 1, while the other categories had a median of 0. To determine whether there was a statistically significant difference in the number of test smells among the subcategories, we performed a Kruskal-Wallis test. As a result, we observed a statistically significant difference ($p < 0.05$). Furthermore, we conducted a post-hoc analysis using Dunn's Post-Hoc Test with a Bonferroni correction to deal with the family-wise error rate. We found that the ``Failures'' and ``Test Completeness'' categories (median=1) contained significantly more test smells than the other three categories (``On-hold Task,'' ``Test Design and Implementation,'' and ``Test Maintenance,'' all median=0), with $p < 0.05$.


\begin{table*}[h]
\small
\centering
\caption{Comparison with the study of Bavota et al.~\cite{DBLP:conf/msr/BavotaR16}}
\label{tab:comparison_bavota}
\begin{tabular}{lll}
\toprule
\textbf{Aspect} & \textbf{Bavota et al. \cite{DBLP:conf/msr/BavotaR16}} & \textbf{Our Study} \\
\midrule
\textit{Studied Ecosystems} & 2 Ecosystems& 24 Ecosystems \\
& {\footnotesize (Eclipse and Apache)}& {\footnotesize (Apache, Spring, Gradle, etc.)}\\
\textit{Studied projects} & 159 projects & 50 projects \\
\textit{Studied year} & 2016& 2025 \\
\textit{Scope of the study} & All Java code   & All Java code \\
& & {\footnotesize (separated into prod. and test code)}\\
\addlinespace
\textit{\# Instances for qualitative analysis} & 366 Prod. SATD & 506 Test SATD\\
\textit{\% Test-related SATD} & 8\% & 15.6\% \\
\textit{\# Test Categories} & 1 category (\ie ``Test'' category) & 5 categories \\
\textit{\# Test Sub-categories} & None & 20 Sub-categories \\
\bottomrule
\end{tabular}
\end{table*}

\subsection{Lessons Learned}\label{sec:lessons}

This study investigates Test SATD to reveal its prevalence, relationship with quality metrics, and types. Additionally, we develop classifiers to categorize the types of Test SATD. Through this comprehensive study, we gained the following five key insights that inform both research and practice in test code quality management. Table~\ref{tab:comparison_bavota} summarizes the differences in findings of the previous study and our study to facilitate our discussion.

\smallskip
\noindent\textbf{Lesson 1. Test SATD Is Distinct from Production SATD.}\\
A key lesson learned is that SATD in test code is not simply a secondary concern to production code SATD; it is a distinct phenomenon with unique characteristics and developer intentions. We identified 20 specific types of issues, categorized into five main groups, many of which do not directly align with existing categories for production code SATD. Notably, test-specific categories like ``Doubts on specific test case design'' emerged from our data but were absent even from general technical debt studies (\ie not only self-admitted technical debt). This divergence demonstrates that assuming similar characteristics for SATD across production and test code is an oversimplification.

Furthermore, we observed a significantly larger proportion of Test SATD (15.6\%) compared to an 8\% finding in a previous study from a decade ago. This temporal shift reflects the increasing emphasis on test writing in modern software development, underscoring the growing importance of addressing test code quality.

\smallskip
\noindent\textbf{Lesson 2. Developers Frequently Leave Incomplete Tests as SATD.}\\
The most frequent category of Test SATD identified was ``Indicates incomplete or unimplemented tests'' (120 out of 430 classified instances). This prevalence indicates a common developer practice of leaving SATD for missing assertions or entire test methods, even when the task might seem minor. A profound lesson here is the need to understand why developers choose to document these incomplete tasks as SATD rather than finishing them immediately. This behavior warrants further qualitative investigation through surveys or interviews to uncover the underlying reasons, such as time constraints, workflow interruptions, or perceived complexity of the remaining task. This insight also suggests a practical application: developers could benefit from specialized code completion tools tailored for test code, which might help address this specific form of technical debt proactively.

\smallskip
\noindent\textbf{Lesson 3. SATD Comments Reveal Issues Beyond Test Smells.}\\
Our study revealed a lack of a direct statistical correlation between Test SATD and test smells. This is a critical lesson: developers are explicitly flagging issues in test code that are often not detected by current automated test smell tools. This implies that existing test smell detection mechanisms are not fully capturing the spectrum of issues developers recognize and admit. This gap presents a clear opportunity for researchers and tool developers to leverage Test SATD comments as a rich, developer-centric source to identify new or overlooked types of test smells, thereby enhancing automated quality assurance tools.

\smallskip
\noindent\textbf{Lesson 4. Manual Classification Is Inherently Ambiguous.}\\
Our manual classification process, while rigorous, highlighted the inherent complexity in precisely categorizing developer intentions behind SATD comments. We employed a card sorting approach with independent labeling by two experienced annotators, and conflicts were resolved by a third, resulting in a Cohen's kappa coefficient of \Kappa, indicating substantial agreement. This experience taught us that even with structured methods and expert annotators, ambiguity and misidentification can occur, suggesting that future efforts in SATD classification could benefit from more context-rich or interactive annotation processes.

\smallskip
\noindent\textbf{Lesson 5. Data Imbalance Hampers ML-Based SATD Classification.}\\
The performance of our machine learning models varied significantly across categories, with a noticeable drop for categories with fewer instances, such as the ``Failures'' category, which had only 15 instances in our dataset. While the CodeBERT-based classifier achieved the highest F1-score of 0.70 overall, its performance for ``Failures'' was the lowest at 0.54. This clearly demonstrates the impact of data imbalance on classification accuracy. A crucial lesson here is the need for a significantly larger and more balanced dataset to improve the robustness and accuracy of automatic SATD classification, especially for less frequent but potentially critical types of debt.

\subsection{Implications}\label{sec:implications}
In this section, we discuss the implications for developers and researchers, mapping the findings to the research questions. 

\smallskip
\noindent
{\bf Implication 1. \ts should not be considered as a negligible concern. } \\ \indent
RQ1 revealed that a non-negligible number of Test SATD instances exist in software repositories. Specifically, software testing plays a crucial role in modern software development, necessitating more proactive test writing than before~\cite{DBLP:journals/access/GurcanDCRS22, DBLP:conf/mipro/HynninenKKT18}.
In fact, we observed a significantly larger ratio of Test SATD compared to a previous study conducted a decade ago~\cite{DBLP:journals/ese/BavotaQOLB15}, despite examining different projects. Many recent studies have reported that the quality of test code leads to serious quality issues in production~\cite{DBLP:journals/tse/AthanasiouNVZ14, DBLP:conf/icsm/SpadiniPZBB18}. This suggests that \textit{researchers should recognize the importance of studying SATD not only in production code but also in test code}. As previous studies have done for production code, future research should investigate the impact of Test SATD on reliability~\cite{DBLP:conf/wcre/WehaibiSG16} and maintainability~\cite{DBLP:conf/icsm/PotdarS14} of software.

\smallskip
\noindent
{\bf Implication 2. Test SATD is likely to be associated with code quality issues. } \\ \indent
RQ2 clarified that methods containing \ts have more lines of code (LOC), annotations, complexity, and code smells than methods without Test SATD. From the perspectives of LOC, annotations, and complexity, this suggests that Test SATD tends to reside in larger and more complex test code. 
However, we also found that Test SATD is not associated with test smells (\ie they are independent). This implies that test smell detection tools are not identifying most of the test issues, which is a gap to be filled by researchers and tool developers. Compared with code smells, test smells are a relatively recent concept and are still under development. Therefore, we recommend that \textit{researchers and tool developers should endeavor to identify new types of test smells, referring to Test SATD in practice.}

\smallskip
\noindent
{\bf Implication 3. Developers should consider using code completion tools specialized in test code.  } \\ \indent
RQ3 categorizes the types of Test SATD. We found that the most frequent category was ``Indicating incomplete or unimplemented tests.'' This suggests that developers often stop writing assertions and document unfinished tests or insufficient test cases as SATD. This behavior may be due to limited time or other constraints, but assertions are usually written with only a few lines of code. It is unclear why developers leave a few-line comments instead of writing assertions. Therefore, we recommend that \textit{researchers investigate why developers did not finish writing assertions through surveys or interviews}. Additionally, \textit{practitioners could consider using copilot tools, especially code completion tools for test code~\cite{DBLP:conf/icse/NieBLMG23,DBLP:conf/iwpc/ZhuLXTZP024}}.


\smallskip
\noindent
{\bf Implication 4. Test SATD can be automatically classified but the performance could be still improved. } \\ \indent
Existing SATD classification tools primarily focused on production code, and no classifier had been developed to categorize SATD in test code. To address this issue, in RQ4, we developed a classifier for \ts and trained it based on the classification results of RQ3. As a result, the classifier utilizing CodeBERT demonstrated the highest performance compared to other models. While the classifiers demonstrated the feasibility of automatically classifying \ts, they achieved a maximum F1-Score of 0.70, which is far from perfect. Therefore, \textit{researchers and tool developers may perform further manual inspections to extend the dataset and retrain the model with a larger dataset to improve the classification performance.}


\section{Threats to validity}
\label{sec:Threat}

\textit{Threats to internal validity} concern the factors we did not
consider that might impact the results. The manual classification in this study was independently conducted by two inspectors to reduce subjective bias. The Cohen's kappa coefficient was 0.78, indicating a high level of agreement. Disagreements were resolved with a third inspector. Despite these measures, the misinterpretation of SATD comments cannot be entirely ruled out. 

\textit{Threats to construct validity} concern the relation between
theory and observation. This study analyzed SATD within test code. A previous study~\cite{DBLP:conf/msr/BavotaR16} has found that SATD instances in the production code may also point out test-related issues, which are not included in our studies. We believe the number of SATD instances related to testing in the production code is relatively small, but future research could consider production code when identifying test-related SATD.

The imbalanced distribution across categories poses validity concerns. The most infrequent category, ``Failures,'' contains only 15 instances, which may not adequately represent the full spectrum of this construct. While this imbalance problem is common in software engineering research~\cite{DBLP:journals/isci/SeiffertKHF14, DBLP:journals/tse/Tantithamthavorn20}, this imbalance directly affects the validity of our performance measurements: during 10-fold cross-validation, several test sets contained only a single ``Failures'' instance, potentially leading to unstable performance estimates that may not accurately reflect the model's true ability to identify this category. While using median performance metrics partially addresses this threat, we cannot claim equal measurement validity across all categories. 

\textit{Threats to external validity} concern the generalizability of our findings. In this study, the manual classification was performed on a sampled subset of detected SATD. To mitigate sampling bias, we applied random sampling to satisfy a 99\% confidence level and a 5\% margin of error, and the study included 50 repositories, which is more than in other studies~\cite{DBLP:journals/infsof/KashiwaNKKSSU22, DBLP:journals/infsof/FariasNKS20}. Nevertheless, the possible bias due to repository selection remains. Furthermore, our work primarily studied open-source projects because code access was necessary for the analysis. It is unclear whether proprietary projects exhibit the same patterns as shown in our results.

In addition, our sample of 506 instances may underrepresent rare but potentially important Test SATD categories (\eg our ``Failures'' category has only 15 instances). However, from a statistical perspective, the distribution of categories in our sample likely reflects their natural prevalence in software systems in the wild. The rarity of certain SATD types, such as ``Failures,'' may indicate their actual low frequency of occurrence rather than a sampling bias. While increasing the dataset size might yield more instances of rare categories, the proportional distribution might remain similar.

    

\section{Conclusion}
\label{sec:Conclusion}

In this study, we conduct a large-scale empirical investigation into the nature of Self-Admitted Technical Debt in test code, an area often overlooked in existing research. 
    We collected 17,766 SATD comments (14,987 from production code, 2,779 from test code) from 50 repositories and analyzed their prevalence, relationship with various quality metrics, specific types, and the feasibility of their automatic classification. 
    
    Our findings reveal several key insights. First, Test SATD is a non-negligible phenomenon, accounting for 15.6\% of all SATD instances across 50 projects. Second, we found that Test SATD correlates with general code quality issues like code smells and complexity, but notably, it does not correlate with test smells, suggesting that SATD and test smells represent distinct quality concerns. Third, through manual analysis, we developed a detailed taxonomy of 20 types of Test SATD, finding that the most common reason developers admit debt is due to incomplete or unimplemented tests. Finally, we demonstrated the feasibility of automatically classifying these SATD types, with a CodeBERT-based model showing the most balanced performance, though we also found that modern LLMs like GPT-4.1 can identify unique instances that other models miss.

    Our future work will focus on two main directions: (i) expanding our manually labeled dataset to improve the performance and robustness of our classification models, and (ii) analyzing the evolution of Test SATD over time to understand its lifecycle and resolution patterns.

\begin{acks}
We gratefully acknowledge the financial support of JSPS for the KAKENHI grants (JP21H03416, JP24K02921), the Bilateral Program grant (JPJSBP120239929), as well as JST for the PRESTO grant (JPMJPR22P3), the ASPIRE grant (JPMJAP2415), and the AIP Accelerated Program (JPMJCR25U7).
\end{acks}

\bibliographystyle{ACM-Reference-Format}
\bibliography{reference.bib}

\end{document}